\documentclass[aps,pra,reprint,superscriptaddress,longbibliography,nofootinbib]{revtex4-2}
\usepackage{amsmath,amssymb,mathtools,bm}
\usepackage{braket}
\usepackage{graphicx}
\usepackage{algpseudocode}
\usepackage{qcircuit}
\usepackage{xcolor}
\newcounter{algorithm}
\renewcommand{\thealgorithm}{\arabic{algorithm}}
\newenvironment{algorithm}{%
  \par\medskip\noindent\refstepcounter{algorithm}%
  \begin{minipage}{\linewidth}%
  \hrule\vspace{0.6ex}%
}{%
  \vspace{0.6ex}\hrule%
  \end{minipage}%
  \par\medskip%
}

\usepackage[caption=false]{subfig}

\definecolor{DSyellow}{RGB}{210,160,0}

\begin{document}

\title{A Recursive Module-Coupling Algorithm for Computing  Low-Energy Eigenstates}
\author{Dihang Sun}
\affiliation{Department of Physics, National University of Singapore, Singapore 117551, Singapore}

\author{Nannan Ma}
\email{e0408695@u.nus.edu}
\affiliation{Department of Physics, National University of Singapore, Singapore 117551, Singapore}
\affiliation{Centre for Quantum Technologies, National University of Singapore, Singapore 117543, Singapore}

\author{Ching Hua Lee}
\affiliation{Department of Physics, National University of Singapore, Singapore 117551, Singapore}

\author{Tianqi Chen}
\affiliation{Institute of Advanced Intelligence and Computing (IAIC), Agency for Science, Technology and Research (A*STAR), 1 Fusionopolis Way, No.~16-16 Connexis, Singapore 138632, Singapore\looseness=-1}
\affiliation{Bioinformatics Institute, Agency for Science, Technology and Research (A*STAR), 30 Biopolis Street, No.~07-01 Matrix, Singapore 138671, Singapore}

\author{Jiangbin Gong}
\email{phygj@nus.edu.sg}
\affiliation{Department of Physics, National University of Singapore, Singapore 117551, Singapore}
\affiliation{Centre for Quantum Technologies, National University of Singapore, Singapore 117543, Singapore}

\date{\today}
\begin{abstract}
Finding the eigenstates of a many-body Hamiltonian is a fundamental challenge in physics and computational science. Since the search space grows exponentially with system size, numerous classical and quantum algorithms have been developed to address this problem. A practical strategy is to identify a physics-informed low-dimensional subspace that effectively accommodates the low-lying eigenstates, thereby reducing the computational complexity. In this paper, we propose a recursive module-coupling algorithm, which iteratively treats a system as a composition of locally-coupled smaller modules, with low-energy subspace estimated successively according to the same recursive structure. Unlike the density matrix renormalization group (DMRG) approach that optimizes a global matrix product state through repeated local sweeps and obtains excited states sequentially, our algorithm constructs a physically tailored variational basis from module eigenstates and obtains several low-energy states on an equal footing, leading to substantial speedups if targeting moderate accuracy.
Our proposed method further leads naturally to a recursive quantum variational algorithm, providing a systematic and modular circuit-construction framework compatible with contemporary gate-based quantum architectures. At each recursive level, block encoders are trained to map logical basis states onto the retained physical subspace, within which a variational circuit is subsequently optimized.
Such a quantum-circuit implementation provides not only a quantum multistate eigensolver, but also a systematic prescription for hierarchically constructing quantum state-preparation circuits. Classical simulations demonstrate the accuracy and efficiency of the proposed method, whereas experiments on IBM quantum processors show that eigenstate preparation with reasonable fidelities is achievable even in the current NISQ era. This work is expected to complement and enable existing quantum algorithms, e.g., by offering fast estimates of the energy gap of a many-body quantum system for quantum annealing and by uploading approximate ground states as a promising starting point for further imaginary time evolution.

\end{abstract}

\maketitle

\section{Introduction}

In many areas of physics research, determining the eigenvalues and eigenstates of a many-body Hamiltonian is a fundamental starting point \cite{lin1993exact,qi2011topological}. Of special physical interest are those with low-lying eigenvalues because they not only determine the lowest-energy configurations but also govern the essential features of physical responses to an external control.  Accurately characterizing these low-energy eigenstates is therefore crucial for obtaining reliable physical predictions. Moreover, many computational problems \cite{mohseni2022ising,lucas2014ising} can be reformulated as eigenvalue problems of appropriately designed Hamiltonians. A notable example is the quadratic unconstrained binary optimization problem \cite{Lucas2014IsingFormulations,Kochenberger2014QUBOSurvey,Glover2022QUBOTutorial}. Clearly then, developing efficient methods for obtaining low-lying eigenstates is not only of theoretical importance, but also highly relevant for practical applications.

Because of the exponential scaling of the associated search space, finding many-body eigenstates on classical computers is computationally challenging in many cases. The emergence of quantum computing provides a new avenue for tackling these problems by possibly exploiting uniquely quantum features such as quantum superpositions and the exponentially large Hilbert space \cite{Feynman1982Simulating,Lloyd1996Universal}. Motivated by these advantages, numerous classical and quantum algorithms have been proposed over the past decades ~\cite{abrams1999quantum,huggins2022unbiasing,dorner2009optimal,zhang2025quantum}.


On a classical computer, tensor-network states are used to represent quantum states and operators as tensor products, thereby significantly reducing memory requirements. For example, matrix product states (MPS) are commonly used to represent one-dimensional quantum states, whereas the DMRG method \cite{White1992DMRG,White1993DMRG,schollwock2005density,Schollwoeck2011DMRGMPS} obtains the ground state by optimizing a global MPS through repeated local sweeps. A quantum anolog \cite{liu2025matrix} of such techniques has also been proposed, in which quantum states are used to represent tensors.
In the current noisy intermediate-scale quantum (NISQ) era \cite{Preskill2018NISQ}, the variational quantum eigensolver (VQE) has become particularly popular \cite{Peruzzo2014VQE,McClean2016VQETheory,Cerezo2021VQAReview}. VQE leverages the expressive power of parameterized quantum circuits while delegating the optimization task to a classical optimizer. Despite many successful applications, the optimization process is often hindered by the well-known barren plateau phenomenon \cite{McClean2018BarrenPlateaus}. Other approaches rely more directly on quantum dynamics. One example is imaginary-time evolution, which gradually projects the system toward low-energy states. However, due to its inherently non-unitary nature, implementing imaginary-time evolution on quantum hardware requires nontrivial encodings and must contend with probabilistic success rates \cite{ma2026penalty,McArdle2019ImaginaryTime,Motta2020QITE}. Another prominent approach is quantum annealing, which depends on maintaining adiabatic evolution in order to reach the desired eigenstate \cite{Kadowaki1998QuantumAnnealing,Farhi2000Adiabatic,Farhi2001AdiabaticAlgorithm,Albash2018AdiabaticReview}. In practice, ensuring adiabaticity throughout the evolution is highly challenging (as the gap function is typically unknown) and plays a crucial role in determining the accuracy of the final result.

Different from those algorithms executed directly in the full exponentially large Hilbert space, a common heuristic strategy is to restrict the eigensolver to an effective subspace that contains the target eigenstates with high probability \cite{McClean2017QSE,Colless2018MolecularSpectra}. Solving the eigenvalue problem within such a reduced low-dimensional space is significantly more tractable for both quantum \cite{nakanishi2019subspace} and classical platforms \cite{zhang2025unified,chen2026tangent}. 

For a typical local many-body lattice Hamiltonian, the underlying geometry naturally permits the system to be decomposed into locally interacting modules coupled through lower-dimensional boundaries \cite{qi2013exact,lee2016exact,gu2016holographic}. For short-range interactions, the intermodule coupling is supported only near these boundaries and is therefore subextensive relative to the bulk Hamiltonian of a sufficiently large module, although it need not be perturbatively weak in absolute magnitude. This locality motivates the expectation that tensor products of the retained low-lying eigenstates of the constituent modules can provide an effective variational subspace for the low-energy eigenstates of the coupled system. For a short-range qubit chain, the same low-energy projection and coupling procedure can be organized into a balanced recursive hierarchy and continued until the desired system size is reached. Conceptually, this construction is guided by two closely related ideas from established renormalization-based approaches. From Wilson's numerical renormalization group, we adopt the strategy of enlarging the system iteratively while restricting its description at each stage to a retained low-energy subspace. From real-space block renormalization, we draw on the interpretation of each recursion level as a spatial coarse-graining step \cite{Kadanoff1966,Wilson1975,Morningstar1994,Morningstar1996}. We repurpose these ideas not to construct an RG flow, but to formulate a finite-size multistate eigensolver and, subsequently, a recursive encoded-subspace algorithm.

Specifically, we first investigate in this work the performance of  a recursive module-coupling based multi-state eigensolver as a classical algorithm.   By treating a many-body system as a composition of smaller modules, we hierarchically scale up the size of a many-body system and systematically search for many-body eigenstates \cite{kadanoff1966scaling,wilson1975renormalization,morningstar1994contractor,morningstar1996contractor}. At each layer of algorithm execution,  the above physics consideration motivates us to construct the subspace directly from products of module states, without the variational optimization required by methods such as DMRG, thus significantly reducing the computational cost. Classical numerical simulations demonstrate the effectiveness and accuracy of this proposed approach. One can then scale up the system size by executing the algorithm again at the next layer.  
As elaborated below, our recursive module-coupling algorithm for computing low-energy many-body eigenstates can assist in estimating the gap between the ground state and the excited states, even at the purely classical computation level.  One direct application of this work is thus to guide the execution of quantum annealing algorithms. Our approach is also compatible with the celebrated MPS-based algorithm. Indeed, the basis states used for coupling two modules can be first obtained from DMRG.  The obtained low-energy eigenstates from our algorithm can be used as a starting state of DMRG as well.  

We then study in this work how the hierarchical structure of our module-coupling-based algorithm admits a systematic and modular quantum-circuit realization compatible with current NISQ hardware. In particular, at each recursive level, trained block encoders represent the retained module subspaces, while a logical variational circuit couples these encoded subspaces and searches for the low-energy states of the merged block. The same encoder--coupler--optimization template is employed at every level, providing a standardized circuit-construction prescription with a well-defined logical interface. Consequently, encoders obtained at one level can be reused directly as elementary modules at the next, allowing larger circuits to be assembled recursively from previously optimized components. Since this construction relies on conventional gate-based circuits, local measurements, and hybrid variational optimization, it is directly compatible with current NISQ processors. Proof-of-principle results obtained on IBM quantum hardware demonstrate that low-energy eigenvalue estimates and approximate eigenstates with reasonable computational-basis distribution fidelity can be obtained despite hardware noise. Our recursive module-coupling algorithm can therefore be transferred systematically to quantum circuits, enabling approximate ground and excited states to be prepared directly on quantum hardware as structured initial states for subsequent refinement procedures, such as imaginary-time evolution.

The main contributions of this work can therefore be summarized as follows. We introduce a physics-informed recursive eigensolver that constructs the low-energy manifold directly from module eigenstates; establish its variational and truncation-error properties under recursive merging; demonstrate a substantial computational advantage in the moderate-accuracy multistate regime; and translate the same hierarchical subspace construction into reusable quantum encoders for direct low-energy-state preparation on gate-based quantum hardware.

This paper is organized as follows. In Sec.~II, we introduce the modular projected eigensolver and its recursive construction. In Sec.~III, we present classical numerical demonstrations on practical physical models and analyze its accuracy, robustness, and scale-up behavior. In Sec.~IV, we characterize the truncation-error scaling and computational performance of the recursive construction, with detailed derivations and additional numerical results provided in the appendices. In Sec.~V, we present a quantum-circuit implementation of our recursive approach, so as to explicitly upload the many-body eigenstates on a quantum computer.   The feasibility of our protocol is shown by using Qiskit simulations and IBM quantum hardware, confirming that our proposed method can be feasibly implemented on present-day quantum processors and can obtain and prepare meaningful low-energy eigenstates from our recursive module-coupling based eigensolver. In Sec.~VI, we summarize the results and discuss possible extensions.

\section{Module-coupling based multi-state eigensolver}
\label{sec:method}

In general, a many-body system is described by a Hamiltonian \(H\), whose energy eigenstates can be written as
\begin{equation}
H\ket{\psi_\mu}=E_\mu\ket{\psi_\mu},
\qquad
E_0\le E_1\le \cdots .
\end{equation}
Here \(\mu\) labels eigenstates of \(H\), with \(\mu=0\) corresponding to the ground state. In most cases, the full energy spectrum is not required. Among all eigenstates, the low-lying states are of particular importance, and we therefore place our main focus on this low-energy sector.

Consider a typical many-body Hamiltonian which can be partitioned into two modules \(A\) ]and \(B\) in the following form: 
\begin{equation}
H=H_A\otimes I_B+I_A\otimes H_B+V,
\label{eq:two-block-hamiltonian}
\end{equation}
where \(H_A\) and \(H_B\) are the respective Hamiltonians of the two modules, \(I_A\) and \(I_B\) are the corresponding identity operators, and \(V\) denotes the inter-module coupling that is local, with an energy scale much less than that of the respective modules.  We can hence assume that \(V\) cannot strongly deform the low-energy subspace generated by the isolated modules, or equivalently, not to induce substantial mixing between this subspace and high-energy module excitations.

The inter-module interaction is taken as
\begin{equation}
V=-\sum_{i,j}J_{ij}'\,O_{A,i}\otimes O_{B,j},
\label{eq:general-coupling}
\end{equation}
where \(O_{A,i}\)(\(O_{B,j}\)) is some local operator on module \(A\)(\(B\)), and \(J_{ij}'\) is the coupling strength between  \(O_{A,i}\) and  \(O_{B,j}\) . Different choices of \(J_{ij}'\) specify different coupling geometries between the two modules. 

The key assumption of our method is that the low-energy eigenstates of the full system have dominant support in the product space formed by the low-energy subspaces of subsystems \(A\) and \(B\). This assumption is most natural for short-range systems, where the inter-module coupling is weak relative to the energy separation between retained and discarded module excitations. It may become less accurate for strong or long-range inter-module coupling, or in systems with a dense low-energy spectrum, where a larger retained dimension \(k\) may be required. Once this physically motivated subspace is constructed, the remaining problem becomes an effective Hermitian eigenproblem whose dimension is controlled by the number of retained local low-energy states.

Let \(\ket{\alpha}_A\) and \(\ket{\beta}_B\) denote the eigenstates of the isolated blocks,
\begin{equation}
H_A\ket{\alpha}_A=E_\alpha^A\ket{\alpha}_A,
\qquad
H_B\ket{\beta}_B=E_\beta^B\ket{\beta}_B,
\label{eq:block-eigenstates}
\end{equation}
ordered by increasing energy. For a chosen truncation level \(k\), we define the tensor-product low-energy subspace
\begin{equation}
\mathcal{V}_k=
\operatorname{span}
\left\{
\ket{\alpha\beta}
=
\ket{\alpha}_A\otimes\ket{\beta}_B
\,\middle|\,
\alpha,\beta=0,\ldots,k-1
\right\}.
\label{eq:truncated-subspace}
\end{equation}
The dimension of \(\mathcal{V}_k\) is \(k^2\), which will be much smaller than the whole Hilbert space. If \(P_k\) is the orthogonal projector onto \(\mathcal{V}_k\), the reduced Hamiltonian is
\begin{equation}
H_{\mathrm{eff}}^{(k)}=P_k H P_k.
\label{eq:effective-hamiltonian}
\end{equation}

Thus the \(k^2 \times k^2\) effective Hamiltonian matrix can be assembled as
\begin{equation}
H_{\mathrm{eff}}^{(k)}
=
\operatorname{diag}
\left(E_\alpha^A+E_\beta^B\right)_{\alpha,\beta<k}
-
\sum_{i,j}
J_{ij}'\,o^{A,i} \otimes o^{B,j}.
\label{eq:assembled-effective-hamiltonian}
\end{equation}
where
\begin{equation}
\begin{aligned}
&\bra{\alpha\beta}
\left(
H_A\otimes I_B+I_A\otimes H_B
\right)
\ket{\alpha'\beta'}=
\left(E_\alpha^A+E_\beta^B\right)
\delta_{\alpha\alpha'}\delta_{\beta\beta'} .
\\
&\bra{\alpha\beta}O_{A,i}\otimes O_{B,j}
\ket{\alpha'\beta'}
=
o^{A,i}_{\alpha\alpha'}o^{B,j}_{\beta\beta'},
\end{aligned}
\label{eq:intra-block-matrix-elements}
\end{equation}
and
\begin{equation}
o^{A,i}_{\alpha\alpha'}=
\bra{\alpha}_A O_{A,i}\ket{\alpha'}_A,
\qquad
o^{B,j}_{\beta\beta'}=
\bra{\beta}_B O_{B,j}\ket{\beta'}_B .
\end{equation}

Since $H_{\mathrm{eff}}$ is just a \(k^2 \times k^2\) matrix, the approximate low-energy spectrum can be easily obtained from
\begin{equation}
H_{\mathrm{eff}}^{(k)}c_\mu^{(k)}
=
E_\mu^{(k)}c_\mu^{(k)} .
\label{eq:reduced-eigenproblem}
\end{equation}
Here $\mu = 0, 1, \dots, k-1$, $E_\mu^{(k)}$ is the eigenvalue of $H_{\mathrm{eff}}$, which can be used to approximate the eigenvalue of $H$. \(c_\mu^{(k)}\) is the eigenvector of $H_{\mathrm{eff}}$ which is represented in the product basis \(\{\ket{\alpha\beta}\}_{\alpha,\beta<k}\), and \(\left[c_\mu^{(k)}\right]_{\alpha\beta}\) is the coefficient of \(\ket{\alpha\beta}\). Thus the approximate eigenstate of $H$ is

\begin{equation}
\ket{\psi_\mu^{(k)}}
=
\sum_{\alpha,\beta<k}
\left[c_\mu^{(k)}\right]_{\alpha\beta}
\ket{\alpha}_A\otimes\ket{\beta}_B .
\label{eq:lifted-state}
\end{equation}

\begin{figure}[t]
    \centering
    \includegraphics[width=\linewidth]{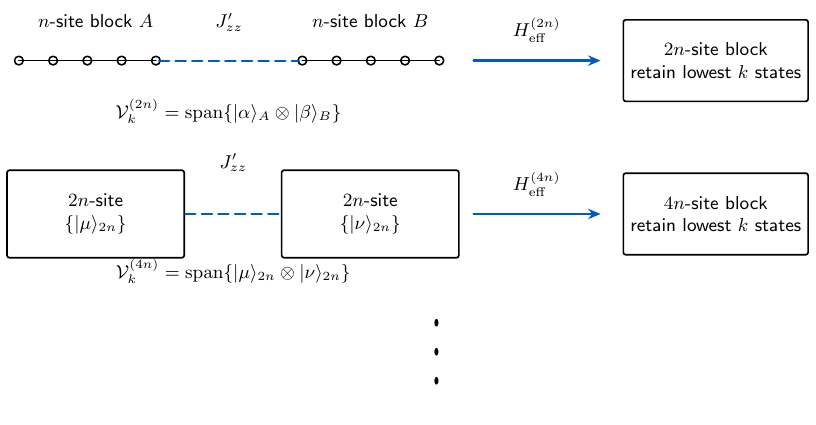}
    \caption{
    Schematic representation of the recursive projection. 
    Two $n$-site blocks $A$ and $B$ are first coupled through the inter-block interaction $J'_{zz}$, giving the projected tensor-product subspace
    $\mathcal{V}^{(2n)}_k=\mathrm{span}\{|\alpha\rangle_A\otimes|\beta\rangle_B\}$.
    Diagonalization of the effective Hamiltonian $H_{\mathrm{eff}}^{(2n)}$ and retention of the lowest $k$ states define the renormalized $2n$-site block. The same merging rule is then applied recursively to construct the $4n$- and $8n$-site block bases from
    $\mathcal{V}^{(4n)}_k$ and $\mathcal{V}^{(8n)}_k$, respectively.
    }
    \label{fig:recursive_block_merging}
\end{figure}

Since we calculate this approximation energy in a subspace, we naturally have 
\begin{equation}
E_\mu^{(k)}\ge E_\mu,
\label{eq:minE}
\end{equation}
And because the subspaces are nested as \(k\) increases, \(\mathcal{V}_k\subset\mathcal{V}_{k+1}\), the Rayleigh-Ritz variational principle gives \cite{Saad2011Eigenvalue}
\begin{equation}
E_\mu^{(k+1)}\le E_\mu^{(k)},
\label{eq:minmax}
\end{equation}
for every eigenvalue index \(\mu\) contained in the truncated space. Thus, for an individual merge with fixed module bases, the computed
energies are variational upper bounds and converge monotonically from
above as \(k\) increases. In the fully recursive construction, the final energies are still the upper bounds. Since changing \(k\) at an earlier level also changes the renormalized module bases entering subsequent levels, the strict monotonicity of the final recursive energies is not guaranteed in general. Nevertheless, increasing \(k\) retains more low-energy information at each merge, so the variational results will still exhibit a clear overall convergence trend toward the reference spectrum.

To process a larger system, the same projection procedure can be applied recursively, as illustrated in Fig.~\ref{fig:recursive_block_merging}. For example, to approximate a 40-site chain, starting from elementary 5-site modules, we first construct an effective problem for a 10-site block. After diagonalizing the corresponding \(k^2\times k^2\) Hamiltonian, the lowest \(k\) eigenvectors are kept as the renormalized basis of the 10-site block. Two such 10-site blocks can then be coupled and projected in the same way, producing a 20-site block basis, and the procedure can be iterated to 40 sites. At each composition step, the diagonalization is performed in a space of dimension \(k^2\), while the physical length of the block doubles. This recursive doubling structure gives the method an intrinsic scale-up capability. For fixed \(k\), each merge involves the same \(k^2\)-dimensional projected problem, while the physical system size doubles at every level. Thus larger systems can be reached by iterating the same controlled projection step, rather than by treating the full Hilbert space directly.

This modular projected eigensolver is summarized in Algorithm~\ref{alg:modular-eigensolver}. The algorithm is stated for a general Hamiltonian decomposed into intra-block Hamiltonians and inter-block couplings.

\begin{algorithm}
\noindent Algorithm~\thealgorithm. Modular projected eigensolver\label{alg:modular-eigensolver}
\vspace{0.6ex}
\begin{algorithmic}[1]
\Require Initial block Hamiltonians \(\{H_b\}\), inter-block couplings \(\{V_{bb'}\}\), subspace dimension \(k\), number of target states \(q\le k\), and target system size.
\Ensure Approximate low-energy eigenpairs \(\{E_\mu^{(k)},\ket{\psi_\mu^{(k)}}\}_{\mu=0}^{q-1}\).
\State Diagonalize each initial block Hamiltonian \(H_b\) and retain the lowest \(k\) eigenstates \(\{\ket{\alpha}_b\}_{\alpha=0}^{k-1}\).
\While{the target system size has not been reached}
    \For{each pair of neighboring blocks \(b,b'\) to be merged}
        \State Form the product basis \(\{\ket{\alpha}_b\otimes\ket{\beta}_{b'}\}_{\alpha,\beta<k}\).
        \State Evaluate all projected coupling matrix elements \(\bra{\alpha\beta}V_{bb'}\ket{\alpha'\beta'}\).
        \State Assemble the \(k^2\times k^2\) effective Hamiltonian \(H_{\mathrm{eff}}^{(k)}\).
        \State Solve \(H_{\mathrm{eff}}^{(k)}c_\mu^{(k)}=E_\mu^{(k)}c_\mu^{(k)}\).
        \State Store the lowest \(k\) eigenvectors through their coefficients in the current product basis, and use the coefficient matrix to transform the projected boundary operators required at the next merge.
        \State Treat these \(k\) vectors and the as the basis of the merged block.
    \EndFor
\EndWhile
\State Return the lowest \(q\) eigenpairs of the final effective Hamiltonian.
\end{algorithmic}
\end{algorithm}

For homogeneous systems, the effective Hamiltonian at each merge has dimension $k^2$, so the time complexity of one merge is $T_{\mathrm{merge}}=\Theta(k^6)$. Since all blocks at a given level are identical and the same renormalized block can be reused, constructing a chain of length $L$ from elementary modules of length $L_0$ requires only $\log_2(L/L_0)$ distinct merge levels. Consequently, once the elementary modules have been prepared, the total merge cost is $\mathcal{O}\!\left(k^6\log_2(L/L_0)\right)$. The complete time and working-memory estimates are summarized in Eqs.~\eqref{eq_merge_complexity} and \eqref{eq_homogeneous_complexity}, with their detailed derivation provided in Appendix~\ref{app:complexity}.

The logarithmic merge structure described above highlights distinction between our modular projected eigensolver and DMRG. Although both methods involve repeated solutions of effective eigenvalue problems \cite{White1992DMRG,White1993DMRG,Schollwoeck2005DMRGReview,Schollwoeck2011DMRGMPS}, their variational spaces are organized in different ways. The density matrix renormalization group represents the full many-body state as a matrix product state and improves it through local tensor updates and repeated sweeps along the chain \cite{Schollwoeck2011DMRGMPS}. Its outer-loop cost therefore scales with the system size, schematically as $\mathcal{O}(L)$ up to sweep and bond-dimension factors. Our method instead constructs the variational space from tensor products of low-energy eigenstates of smaller modules. Starting from blocks of length $L_0$, the recursive construction reaches a chain of length $L$ through only $\log_2(L/L_0)$ merge levels.

The efficiency of this recursive construction comes from the fact that the low-energy subspaces of isolated modules already capture the dominant components of the low-lying states in the coupled system. The resulting variational subspace is therefore much smaller than the full Hilbert space while remaining physically adapted to the problem, so high fidelities can often be achieved with relatively small $k$. By contrast, the density matrix renormalization group typically starts from a low-entanglement matrix product state, often a product state chosen for stability and symmetry control. Its variational manifold is not explicitly constructed from the low-energy eigenstates of the target modules and may therefore require a larger bond dimension and repeated sweeps to reach comparable accuracy.

In addition, in our method, low-energy excited states are obtained simultaneously with the ground state in the same projected eigenproblem at each merge. This differs from excited-state density matrix renormalization group, where one usually needs to enforce orthogonality to previously obtained states, which can slow down convergence. We do not claim that this method is generally superior to the density matrix renormalization group for high-precision one-dimensional ground-state calculations. Rather, its advantage is that it can provide fast, high-quality approximations to a ground state and a few low-energy excited states using a compact, physically tailored subspace. This makes it useful, for example, as an initial-state preparation method for imaginary-time evolution or subsequent dynamical simulations.

\begin{figure*}
\centering
\includegraphics[width=0.98\textwidth]{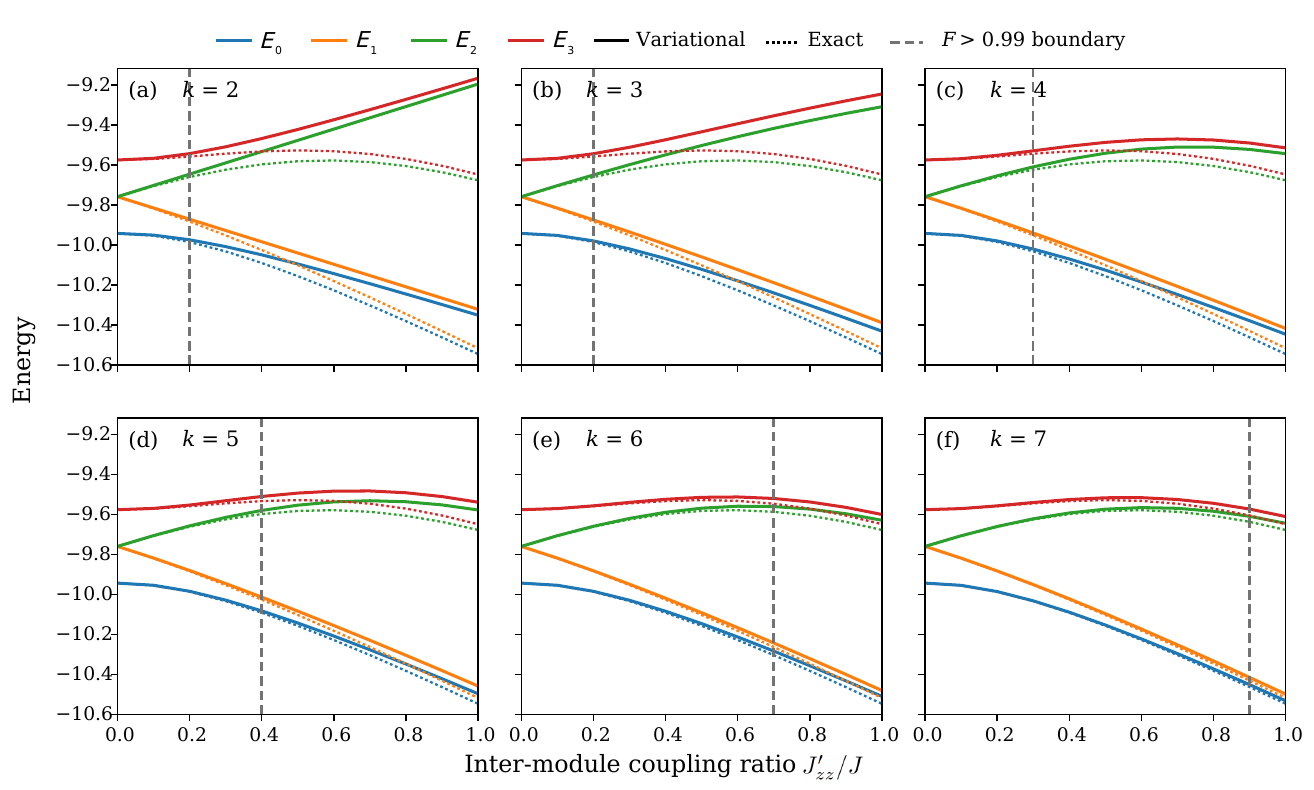}
\caption{
Lowest four energy levels for different subspace dimensions \(k\), obtained at \(J=1.0\) and \(h=0.7\).
In each panel, the solid curves denote the variational energies obtained in the truncated subspace, while the dotted curves denote the exact energies of the full system.
The vertical dashed line marks the boundary of the fidelity criterion \(F>0.99\); namely, the region to the left of the dashed line corresponds to the parameter range in which the variational low-energy states have fidelity greater than \(0.99\).
The variational energies form upper bounds to the exact levels and converge toward the exact spectrum as \(k\) increases.
}
\label{fig:ten-site-spectrum}
\end{figure*}

\section{Numerical Validation of the recursive eigensolver}
\label{sec:classical-results}

In this section, we apply Algorithm~\ref{alg:modular-eigensolver} on a classical computer to confirm the generality and accuracy of the modular projected eigensolver introduced above. In particular, we focus on three aspects. First, we show in a ten-site transverse-field Ising chain that a finite retained dimension \(k\) is sufficient to reproduce the low-energy spectrum accurately, and that the approximation systematically improves as \(k\) increases. Second, we examine different inter-module coupling geometries and weak longer-range coupling terms, demonstrating that the method remains robust under changes in the coupling structure. Third, we apply the recursive construction to longer chains in a practical physical model and analyze how the truncation error behaves with system size. The results show that the error accumulation is well behaved, with an approximately linear growth of the total ground-state energy error and a controlled error density in the tested regime.

\subsection{Ten-site model}
\label{subsec:ten-site-benchmark}

We first test one iteration of the projected eigensolver on a ten-site transverse-field Ising chain consisting of two five-site modules, where each module can be described by the Hamiltonian 
\begin{equation}
H_{A(B)}=-J\sum_{i=1}^{L-1}\sigma_i^z\sigma_{i+1}^z-h\sum_{i=1}^{L}\sigma_i^x .
\label{eq:block-hamiltonian}
\end{equation}
Here $L$ denotes the length of the chain, \(J\) is the intra-module Ising coupling strength and \(h\) is the transverse-field strength. In this section, we use the nearest-neighbor inter-module coupling
\begin{equation}
V=-J_{zz}'\,\sigma_{A,5}^z\otimes\sigma_{B,1}^z,
\label{eq:nearest-coupling}
\end{equation}
with $J_{zz}'$ ranging from $0$ to $1$. The dependence on the coupling position
and coupling type is analyzed in Appendix~\ref{app:coupling-geometry}. Exact diagonalization of the full ten-spin Hamiltonian is used as the reference. Figure~\ref{fig:ten-site-spectrum} compares the lowest four energy levels obtained from the projected method with the exact spectrum for several truncation levels $k$.
\begin{figure*}
\centering
\includegraphics[width=0.90\textwidth]{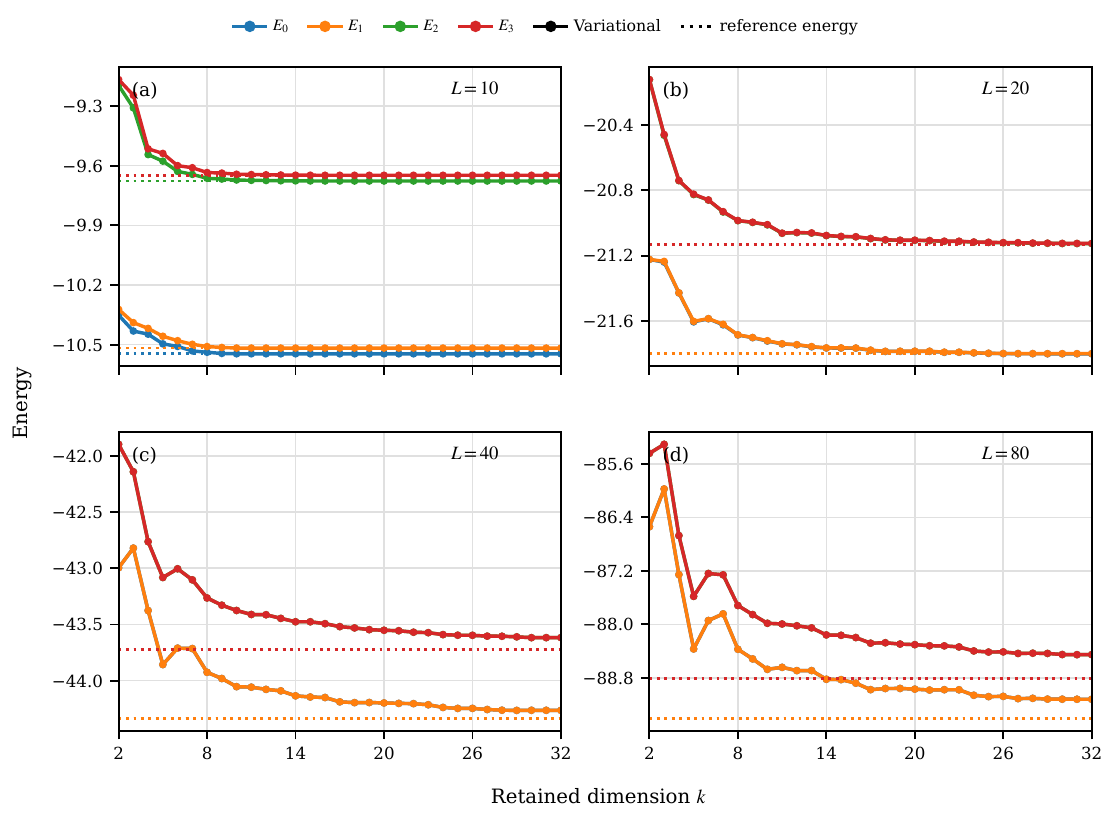}
\caption{
Low-energy spectra for longer transverse-field Ising chains obtained at \(J=1.0\) and \(h=0.7\).
The four panels correspond to \(L=10,20,40,80\), respectively.
Solid curves denote the low-energy levels obtained from the recursive truncated-subspace construction, while dotted horizontal lines denote the corresponding DMRG reference energies.
The results show that increasing the subspace dimension \(k\) improves the low-energy spectrum, with the convergence being more rapid for shorter chains and gradually more demanding as the system size increases.
}
\label{fig:long-chain-spectrum}
\end{figure*}

\begin{figure}
\centering
\includegraphics[width=0.9\columnwidth]{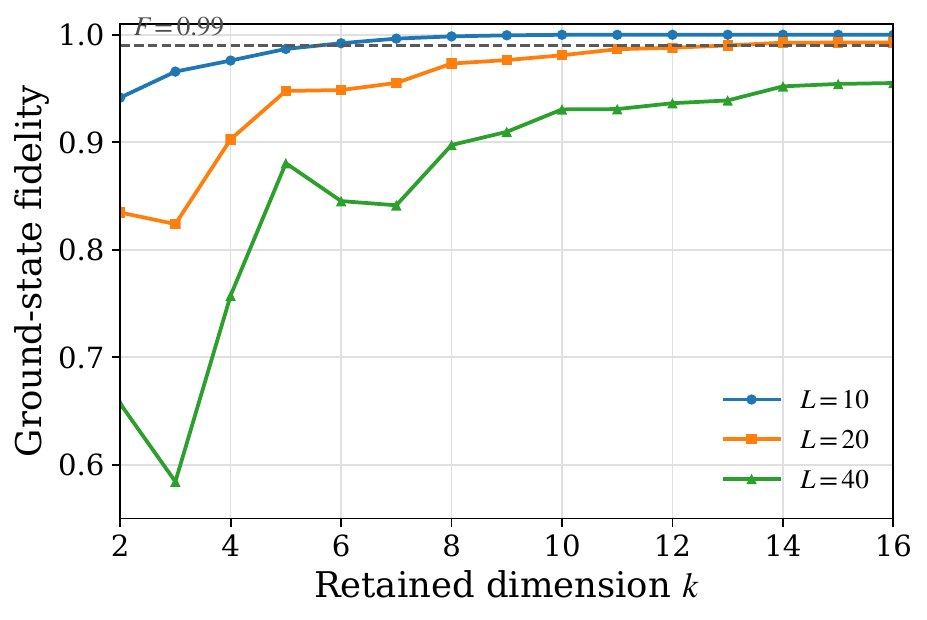}
\caption{
Ground-state fidelity as a function of the subspace dimension \(k\) for longer transverse-field Ising chains, obtained at \(J=1.0\) and \(h=0.7\).
The three curves correspond to system sizes \(L=10\), \(L=20\), and \(L=40\), respectively.
The horizontal dashed line marks the reference value \(F=0.99\).}
\label{fig:ground-state-fidelity-long-chain}
\end{figure}

In Fig.~\ref{fig:ten-site-spectrum}, the variational levels remain above the exact levels and move downward as $k$ increases, in agreement with Eq.~\eqref{eq:minmax}. For small $k$, the truncated space captures the qualitative low-energy band structure but loses quantitative accuracy as the coupling becomes stronger. For $k\ge 6$, the projected and exact spectra become nearly indistinguishable on the scale shown in Fig.~\ref{fig:ten-site-spectrum}. Quantitatively, the full Hilbert space has dimension $2^{10}=1024$, while the reduced space has dimension $k^2$. Thus $k=4$ uses only $16$ states, about $1.6\%$ of the full space, and $k=7$ uses $49$ states, about $4.8\%$ of the full space. 

When an exact reference state \(\ket{\psi_\mu}\) is available, we can also evaluate the state fidelity
\begin{equation}
F_\mu^{(k)}
=
\frac{
\left|\braket{\psi_\mu|\psi_\mu^{(k)}}\right|^2
}
{
\braket{\psi_\mu|\psi_\mu}
\braket{\psi_\mu^{(k)}|\psi_\mu^{(k)}}
}.
\label{eq:fidelity}
\end{equation}
For \(k=7\), the fidelity of the considered low-energy states remains above \(0.99\) throughout the entire range shown in Fig.~\ref{fig:ten-site-spectrum}. The results show that a small product subspace constructed from the isolated-module low-energy eigenstates is sufficient to reproduce the low-energy band structure over a broad range of inter-module coupling strengths.

\subsection{Scale-up behavior in practical physical model}
\label{subsec:longer-chains}

\begin{figure*}
\centering
\includegraphics[width=0.9\textwidth]{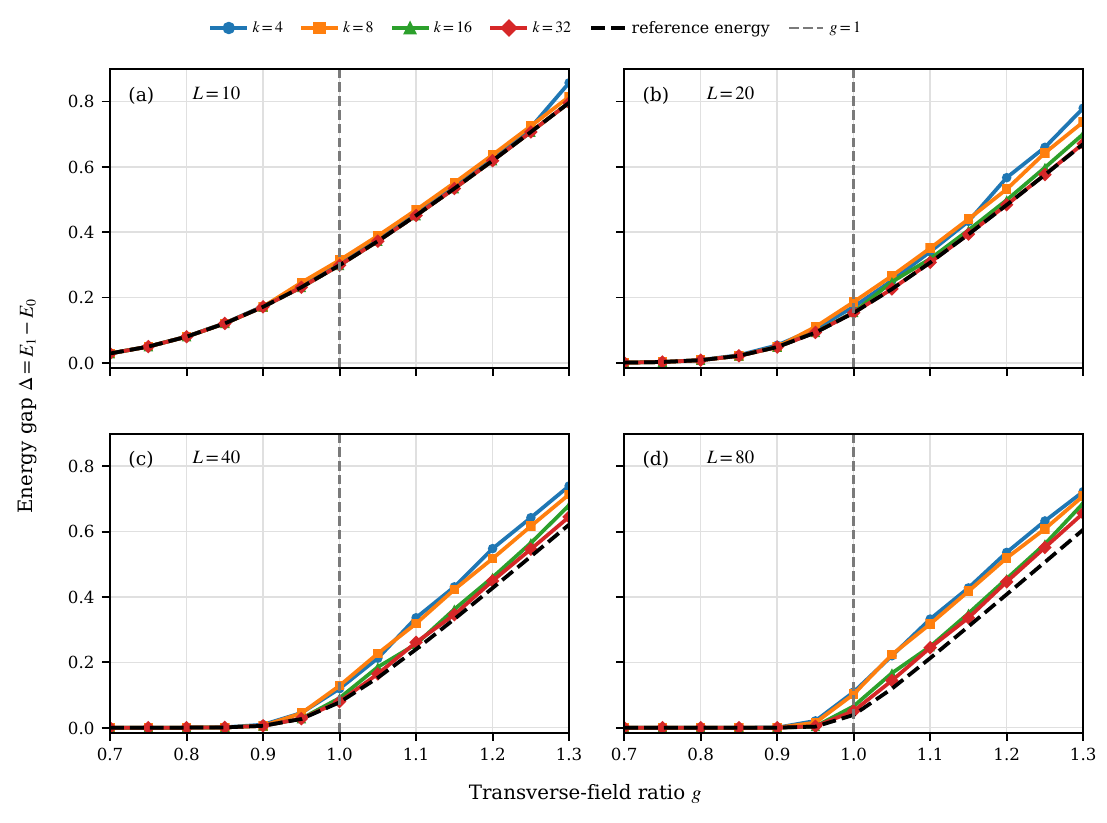}
\caption{
Energy gap \(\Delta=E_1-E_0\) as a function of the transverse-field ratio \(g\), obtained at \(J=1.0\).
The four panels correspond to \(L=10,20,40,80\), respectively.
For each system size, the solid curves show the results obtained from the recursive truncated-subspace construction with \(k=4,8,16,32\), while the black dashed curve denotes the DMRG reference result.
The vertical dashed line marks the critical point \(g=1\).
}
\label{fig:phase-transition-gap}
\end{figure*}

Having validated our method on $L=10$ chains (which is a single merge of two five-site modules), we now test it in a larger physical model of practical relevance, where reaching the target system sizes requires the recursive construction. This allows us to examine how the approximation scales under repeated merges and how the truncation error accumulates with system size. We use the open-boundary transverse-field Ising Hamiltonian
\begin{equation}
H=-J\sum_{i=1}^{L-1}\sigma_i^z\sigma_{i+1}^z-h\sum_{i=1}^{L}\sigma_i^x,
\label{eq:long-chain-tfim}
\end{equation}
where the inter-block coupling is taken to be the same nearest-neighbor Ising interaction as the intra-block coupling. Starting from $L=5$ blocks, we use the recursive method to construct $L=10$, $L=20$, $L=40$, and $L=80$ chains. At each stage, the calculation keeps the lowest \(k\) states of the current block, and uses their tensor products to form the variational space for the next merge. Therefore, the effective dimension of each merge problem is fixed by \(k^2\). Since $L$ is large, the exact diagonalization is not applicable, here we use the DMRG results as a reference, which were obtained with the two-site algorithm implemented in TeNPy~\cite{Hauschild2018,Hauschild2024,Hauschild2024codebase}, at a
maximum bond dimension $\chi_{\max}=128$ throughout. Sweeps were terminated once the energy and entanglement entropy changed by less than $10^{-8}$ and $10^{-5}$, respectively, or after eight sweeps, with the discarded weight per
bond capped at $10^{-8}$ and singular values below $10^{-10}$ truncated.

The spectra of the lowest four energy levels for different system sizes are shown in Fig.~\ref{fig:long-chain-spectrum}. For $L=10$ and $L=20$, the projected spectra closely match the reference values. For $L=40$ and $L=80$, the approximate energies become slightly higher, as expected from the variational character of the truncation. The deviation reflects the fact that a fixed subspace dimension $k$ becomes more restrictive as the chain length grows.

The ground-state fidelity as a function of the subspace dimension \(k\) is shown in Fig.~\ref{fig:ground-state-fidelity-long-chain}. Although the fidelity decreases as the system size \(L\) increases, high fidelity can still be obtained with a relatively modest retained dimension. In particular, the required value of \(k\) remains small compared with the dimension of the full Hilbert space, indicating that the low-energy modular basis continues to provide a compact representation for longer chains.

We further test whether the same low-energy description captures the qualitative gap structure near the phase transition point \cite{Pfeuty1970TFIM,Sachdev2011QPT}. Defining $g=h/J$, the Jordan-Wigner solution of the infinite chain gives the quasiparticle dispersion \cite{Jordan1928Wigner,Lieb1961SolubleModels,Pfeuty1970TFIM}
\begin{align}
\varepsilon(q)=2J\sqrt{1+g^2-2g\cos q},
\\
\Delta_{\mathrm{bulk}}=\min_q\varepsilon(q)=2J|1-g|.
\label{eq:jw-dispersion}
\end{align}
For an open finite chain in the ferromagnetic regime $g<1$, the two nearly degenerate lowest states are associated with the exponentially small splitting between the two boundary-Majorana sectors \cite{Kitaev2001Majorana},
\begin{equation}
\Delta_0(L)=E_1-E_0\sim e^{-L/\xi(g)}.
\label{eq:edge-splitting}
\end{equation}
For $g>1$, the chain is in the paramagnetic regime and the lowest excitation is a bulk Bogoliubov quasiparticle, so that the first excitation gap increases approximately as
\begin{equation}
\Delta_{\mathrm{bulk}}=E_1-E_0\approx 2J(g-1).
\label{eq:paramagnetic-gap}
\end{equation}
The finite-size gap therefore gives a qualitative test of whether the truncated basis preserves the low-energy structure across the transition region.

The data in Fig.~\ref{fig:phase-transition-gap} show that the projected description can qualitatively keep the expected finite-size signature of the transition. In the ordered regime, the ground and first excited states remain nearly degenerate for sufficiently long chains. In the disordered regime, the lowest gap opens and follows the bulk excitation scale. This indicates that the retained low-energy block states preserve not only individual energies but also the qualitative spectral rearrangement associated with the phase transition.

Overall, the numerical results support the validity of our proposed recursive module-coupling-based eigensolver as an efficient scheme for obtaining low-energy eigenstates. The short-chain tests show that a small fraction of the full Hilbert space is sufficient to reproduce the low-energy spectrum with high accuracy, and the same recursive construction can be extended to longer chains, although a fixed $k$ becomes more restrictive as the system size increases. Our results also show that the method preserves the qualitative gap structure across the ferromagnetic and paramagnetic regimes. Taken together, these results demonstrate that the retained low-energy module states provide a physically meaningful truncated basis for constructing the low-energy subspace of the composed many-body system at the next hierarchical level.

\section{Truncation error bounds}
\label{sec:error-bounds}

\begin{figure*}
\centering
\includegraphics[width=0.9\textwidth]{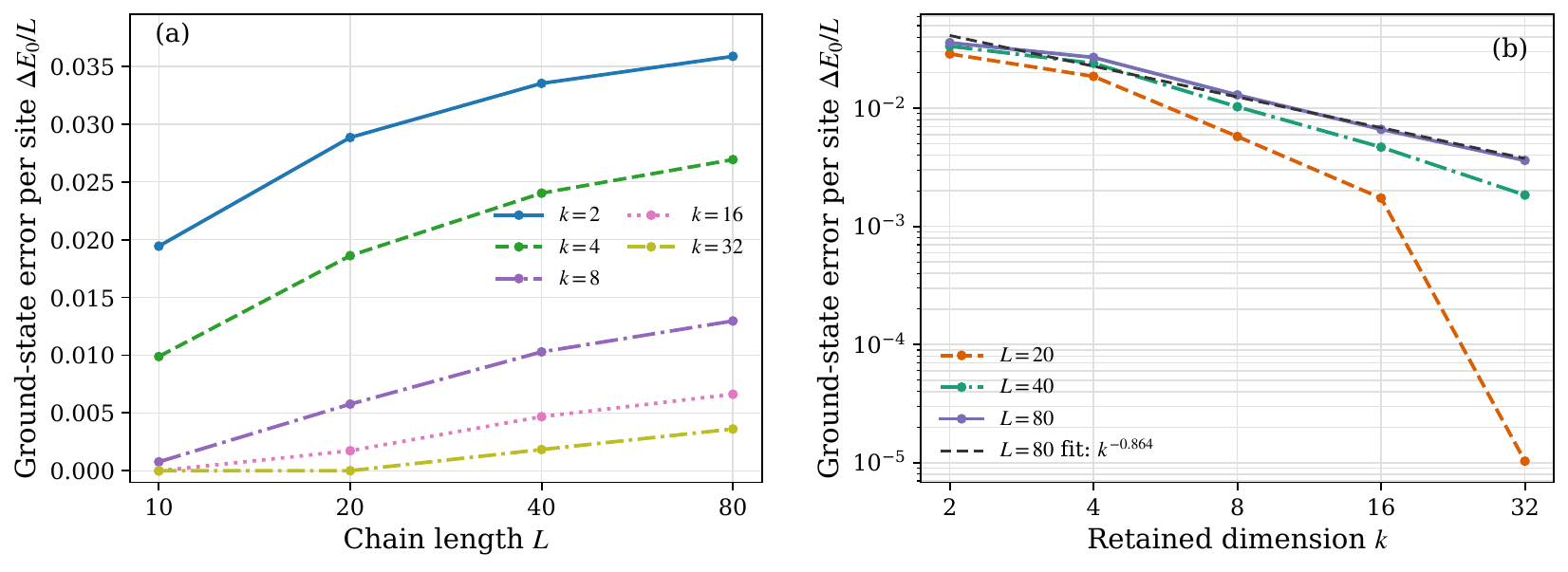}
\caption{
Ground-state energy error per site for different numbers of retained dimensions per iteration.
The error is defined by
$\Delta E_0=E_0^{(k)}-E_0^{\mathrm{ref}}$.
All points at $L=10,20,40,80$ are obtained directly from the modular projected eigensolver at $J=1.0$ and $g=0.7$.
Panel (a) shows that the error density approaches an approximately size-independent bound, while panel (b) shows the systematic decrease with $k$.
The dashed line shows a finite-range empirical fit to the $L=80$ data.
}
\label{fig:error-scaling}
\end{figure*}

To analyze how the truncation error accumulates under recursive merging, we study the dependence of the ground-state energy error on the system size for different truncation levels $k$. We define
\begin{equation}
\Delta E_0(L,k)
=
E_0^{(k)}(L)-E_0^{\mathrm{ref}}(L),
\label{eq:ground-state-error}
\end{equation}
where $E_0^{(k)}(L)$ is the ground-state energy obtained from the modular projected eigensolver and $E_0^{\mathrm{ref}}(L)$ is the exact free-fermion reference energy. Since the total energy is extensive, it is natural to examine the error per site, $\Delta E_0/L$, which is plotted in Fig.~\ref{fig:error-scaling} for $k=2,4,8,16,32$.

From this figure, we can find that the truncation error remains controlled throughout the recursive construction. Let $P=P_k$ project onto the retained product sector and let $Q=I-P$. For a single merge, write $A=PHP$, let $a=\lambda_{\min}(A)$ be the projected ground-state energy, let $b=\lVert PHQ\rVert_2$ quantify the coupling between the retained and discarded sectors, and let $\gamma>0$ denote their energy separation. The error introduced by that merge is bounded by
\begin{equation}
0\leq a-E_0^{\mathrm{ref}}(L)
\leq
\frac{\sqrt{\gamma^2+4b^2}-\gamma}{2}
\leq
\frac{b^2}{\gamma}.
\label{eq:single-merge-bound-main}
\end{equation}
Under the assumed nonzero discarded-sector separation, $b^2/\gamma$ remains finite throughout the recursive construction. Assuming that the single-merge error is uniformly bounded by $R(k)$, propagating this result through our recursive construction shows that
\begin{equation}
\frac{\Delta E_0(L,k)}{L}
\leq
\frac{R(k)}{L_0}
\left(1-\frac{L_0}{L}\right)
<
\frac{R(k)}{L_0}.
\label{eq:error-density-bound-main}
\end{equation}
Thus, the total truncation error grows at most extensively with $L$, while the error density remains bounded. The complete derivation of Eqs.~\eqref{eq:single-merge-bound-main} and \eqref{eq:error-density-bound-main} is provided in Appendix~\ref{app:error-bounds}. The direct numerical validation for $L=10$ and $L=20$, shown in Fig.~\ref{fig:bound-validation} of that appendix, confirms that the actual error remains below the nonperturbative bound for every tested $k$.

The dependence on $k$ can be understood from the spectral weight removed at each merge. Increasing $k$ moves the discarded-sector cutoff to higher excitation energies and includes more of the boundary-coupled spectral weight in the retained subspace, thereby increasing the calculation accuracy. The corresponding analysis is also given in Appendix~\ref{app:error-bounds}. A log--log fit of the $L=80$ data for $k=4,8,16,32$ in Fig.~\ref{fig:error-scaling} gives
\begin{equation}
\frac{\Delta E_0(L,k)}{L}\sim k^{-0.87},
\end{equation}
showing that the error per site follows a near-inverse dependence on $k$ over the fitted range. The precise exponent is model dependent because it is determined by the excitation spectrum and the boundary-operator matrix elements.

We also verify that the qualitative size dependence is robust against the choice of the elementary module length. The results for $L_0=5$, $7$, and $9$, presented in Fig.~\ref{fig_L0_scaling} of Appendix~\ref{app:error-bounds}, show that for each fixed $L_0$ and $k$, the error density approaches an approximately size-independent value, while increasing $k$ systematically reduces the error for every module size.

These results also characterize the trade-off between accuracy and computational cost. Over the relevant finite range of $k$, writing $R(k)\simeq C_Rk^{-p}$ and using the merge cost $T_k\simeq c_Tk^6$ gives
\begin{equation}
\frac{\Delta E_0(L,k)}{L}
=
\mathcal{O}\!\left(T_k^{-p/6}\right),
\label{eq_cost_error_tradeoff}
\end{equation}
as derived in Appendix~\ref{app:error-bounds}. For the fitted value $p\simeq0.87$, the error density therefore scales approximately as $T_k^{-0.145}$. For example, doubling $k$ increases the merge cost by a factor of $2^6=64$, while reducing the estimated error density to $2^{-0.87}\simeq0.55$ of its original value.

\begin{figure}[t]
\centering
\includegraphics[width=\linewidth]{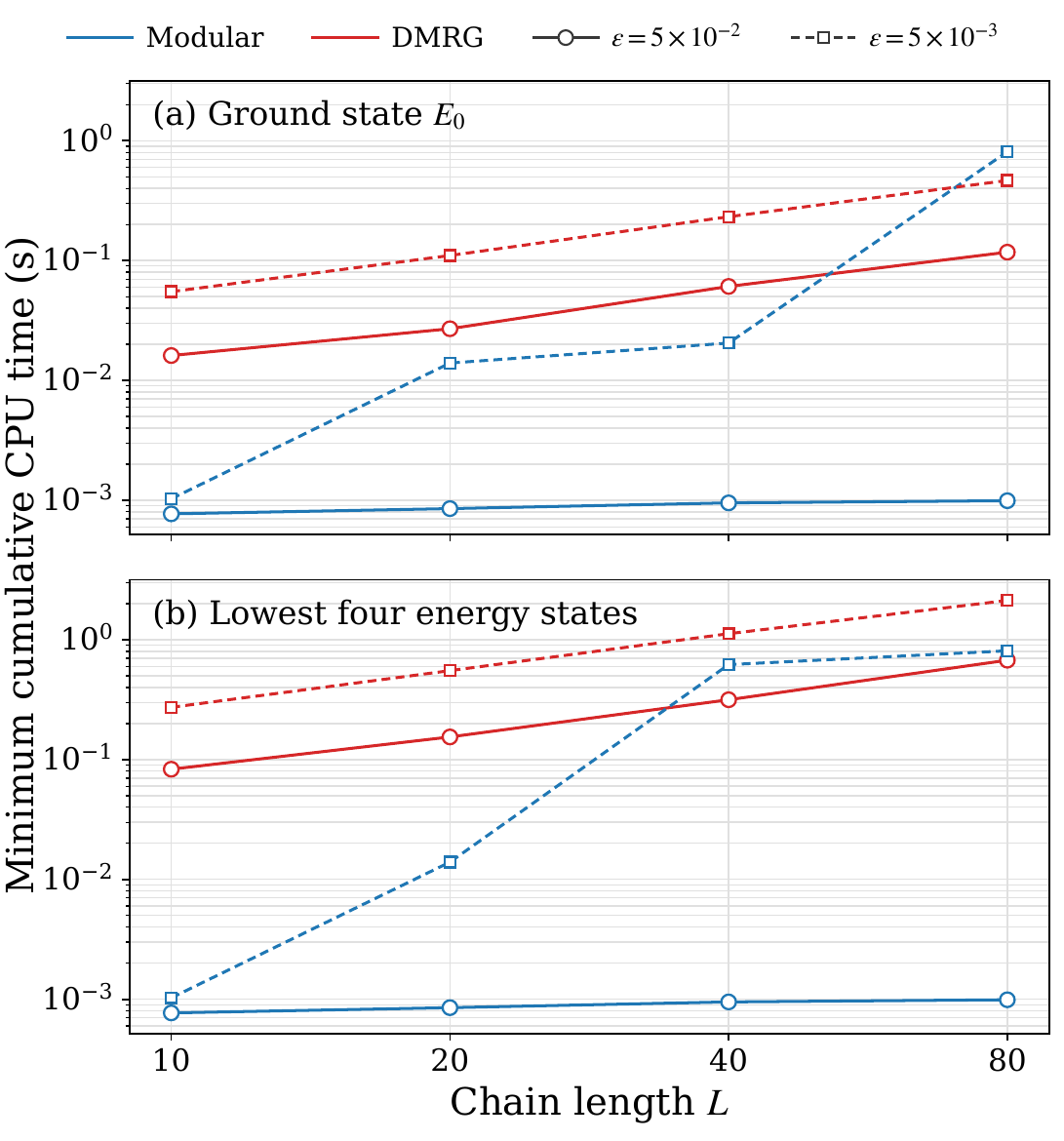}
\caption{
Comparison of the minimum cumulative CPU time required by the modular projected eigensolver and DMRG to reach fixed error-density targets for $J=1.0$, $g=0.7$, and $L_0=5$.
Panel (a) considers the ground-state energy alone, while panel (b) requires all four lowest energies $E_0,E_1,E_2,E_3$ to satisfy the stated error-density threshold.
Solid curves with circles correspond to
$\epsilon_{\mathrm{tar}}=5\times10^{-2}$, and dashed curves with squares correspond to
$\epsilon_{\mathrm{tar}}=5\times10^{-3}$.
All runtime benchmarks were performed using a single CPU thread on an AMD EPYC 9V74 80-Core Processor.
}
\label{fig:runtime_comparison}
\end{figure}

To benchmark this finite-range cost--accuracy trade-off, we compare the modular projected eigensolver with DMRG at fixed error-density thresholds. For each method, we define the error density of the $j$th energy level as
\( \epsilon_j(L) = \left|E_j(L)-E_j^{\mathrm{ref}}(L)\right|/L, \)
where $E_j^{\mathrm{ref}}(L)$ is the exact free-fermion reference energy. In Fig.~\ref{fig:runtime_comparison}(a), the target condition is imposed only on the ground state,
$\epsilon_0\leq\epsilon_{\mathrm{tar}}$. In Fig.~\ref{fig:runtime_comparison}(b), all four lowest levels are required to satisfy the target simultaneously, \(\max_{0\leq j\leq3}\epsilon_j
\leq \epsilon_{\mathrm{tar}} \).

At the relaxed threshold
$\epsilon_{\mathrm{tar}}=5\times10^{-2}$, corresponding to relative errors of approximately $4$--$5\%$, the modular method is tens of times faster than DMRG throughout the tested range of $L$ for both tasks. Tightening the threshold to
$5\times10^{-3}$, corresponding to relative errors of approximately $0.4$--$0.5\%$, requires larger retained dimensions and therefore increases the merge cost, narrowing the computational advantage.

Notably, the modular method is especially efficient when the four lowest levels are calculated simultaneously, because all four levels are obtained from a single projected eigendecomposition, whereas DMRG calculates the excited states sequentially. These results directly illustrate the cost--accuracy trade-off in Eq.~\eqref{eq_cost_error_tradeoff} and show that when moderate accuracy is sufficient, the modular construction offers a substantial computational advantage over DMRG, particularly when several low-energy states are required.

Overall, the theoretical and numerical results show that the total truncation error grows at most extensively with $L$, while the error density approaches an approximately size-independent value, and decreases systematically with the retained dimension $k$. Together with the running-time comparison, these findings provide a coherent theoretical and numerical characterization of the accuracy and efficiency of our modular construction method. The complete mathematical derivations and additional numerical tests are collected in Appendix~\ref{app:error-bounds}.

\section{Encoded quantum-circuit implementation}
\label{sec:hardware}

Although the classical modular projection greatly reduces the computational cost and can achieve high fidelity with a compact retained subspace, explicitly reconstructing the retained basis states in the full physical Hilbert space becomes prohibitive as the system size \(L\) increases. Even when these states are represented implicitly through recursive projection maps, contracting the resulting representations to evaluate projected matrix elements or physical observables can remain computationally demanding for large blocks or large retained dimensions. Moreover, in many applications, obtaining the eigenvalues alone is insufficient; the corresponding low-energy states must also be prepared for subsequent measurements or further quantum-information processing. These considerations motivate a quantum-circuit formulation of the modular projected eigensolver. Importantly, the recursive module structure in our approach translates naturally into a modular and standardized circuit-construction rule. This uniform encoder--coupler architecture enables larger circuits to be assembled hierarchically from previously optimized components, without redesigning a full system-scale ansatz from scratch, thereby providing an efficient quantum realization of the modular projected eigensolver.

In the classical approach, the retained low-energy eigenstates of each module are represented as amplitude vectors, and the effective Hamiltonian is assembled from their matrix elements. In the circuit formulation, we instead introduce a modular encoded-subspace eigensolver in which block encoders are trained to map logical basis states to the corresponding physical states in the retained low-energy subspace~\cite{Cerezo2021VQAReview,Kandala2017HardwareEfficient,romero2017quantum}. Here the logical basis is the computational basis of a small logical register, whose basis states label the retained low-energy states. For example, if \(k\) states are retained, one may use \(m=\lceil \log_2 k\rceil\) logical qubits and identify the first \(k\) computational-basis states with the retained-subspace labels. This replaces explicit wave-function storage with a representation in terms of trained circuit parameters.

The principal purpose of this encoding is to realize the retained low-energy subspace as a family of physically preparable states and to enable the subsequent variational search within the corresponding logical subspace. A related quantum MPS approach represents individual MPS tensors as quantum states and implements a quantum analogue of single-site DMRG through repeated local VQE updates, variational quantum SVD, quantum reshape operations, and back-and-forth sweeps~\cite{liu2025matrix}. However, these repeated variational circuits introduce additional optimization and measurement cost, as well as potential error accumulation. By contrast, our method directly encodes physically tailored module low-energy subspaces and recursively reuses the trained encoders, therefore substantially reducing the associated circuit complexity.

The quantum implementation provides two routes to the low-energy eigenproblem. One may prepare encoded logical basis states and measure the Pauli terms of the physical Hamiltonian to reconstruct the matrix elements of the effective Hamiltonian, thereby assisting the classical projected eigensolver. Alternatively, the eigenproblem can be processed directly on the quantum platform. After the block encoders have been trained and fixed, one may optimize a logical variational circuit directly within the encoded subspace to prepare the ground state and low-energy excited states, without explicitly constructing the effective Hamiltonian. The quantum implementation therefore connects the truncated subspace and physical state preparation within a unified circuit framework. The purpose of this section is to describe this circuit-level realization.

\subsection{Circuit-level subspace algorithm}
\label{subsec:block-isometry}

Having established this motivation, we now describe how the projected eigensolver is realized as a trainable circuit. The circuit construction separates the task into two variational stages. First, each module is assigned a parameterized encoder whose image represents the retained local low-energy subspace. After these local encoders are trained and fixed, they are combined into a global encoder, and a logical unitary is optimized within the resulting tensor-product encoded space.

\begin{figure}[t]
\centering
\centering
\includegraphics[width=0.98\columnwidth]{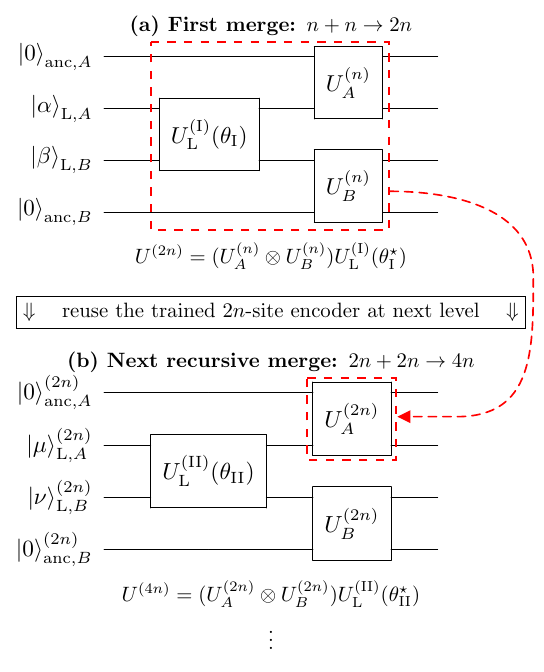}
\caption{
Circuit structure of the encoded-subspace algorithm.
(a) The logical unitary \(U_{\mathrm{L}}(\theta)\) acts on the logical input states \(\ket{\alpha}_{\mathrm{L},A}\) and \(\ket{\beta}_{\mathrm{L},B}\). The two trained block encoders \(U_A\) and \(U_B\) then map the logical degrees of freedom together with the ancillas into the physical encoded subspace. (b) shows the next recursive
level, where two copies of the trained \(2n\)-site encoder are combined
with a new logical unitary \(U_{\mathrm{L}}^{(\mathrm{II})}(\theta_\mathrm{II})\) to construct
the \(4n\)-site encoder. The same procedure can be applied recursively.
}
\label{fig:encoded-circuit-overview}
\end{figure}

The circuit structure is summarized in Fig.~\ref{fig:encoded-circuit-overview}.  In the first stage, each local module is assigned a block encoder. For a block \(A\) of \(n_A\) physical qubits, let \(k\) be the desired retained subspace dimension and define
\begin{equation}
m_A=\left\lceil \log_2 k\right\rceil .
\end{equation}
The first \(m_A\) qubits represent the logical label \(\alpha\), while the remaining \(n_A-m_A\) qubits are initialized as ancillas. A parameterized unitary \(U_A\) is trained so that the first \(k\) logical inputs are mapped to a low-energy subspace $\{\ket{\alpha}_A\}$ of the block Hamiltonian ,
\begin{equation}
U_A
\left(
\ket{\alpha}_{\mathrm{L}}\otimes
\ket{0}^{\otimes(n_A-m_A)}
\right)
\approx
\ket{\alpha}_A,
\qquad
\alpha=0,\ldots,k-1 .
\label{eq:block-encoder}
\end{equation}
Equivalently, after training, the block isometry is implemented as
\begin{equation}
W_A
=
U_A
\left(
I_{2^{m_A}}\otimes
\ket{0}^{\otimes(n_A-m_A)}
\right).
\label{eq:isometry-from-unitary}
\end{equation}
For multiple blocks, the fixed global encoder is
\begin{equation}
U_{\mathrm{enc}}
=
\bigotimes_b U_b,
\qquad
W=
\bigotimes_b W_b .
\end{equation}

In the second stage, the trained block encoders are held fixed. We then introduce a new logical circuit \(U_{\mathrm{L}}(\theta)\), which acts only on the retained logical qubits. Since the block encoders map the logical basis into the physical low-energy subspace, variational operations on the logical qubits are equivalent to variational operations within the low-energy subspace of the full system. The encoded variational state has the form
\begin{equation}
\ket{\Psi(\theta)}
=
U_{\mathrm{enc}}
\left(
U_{\mathrm{L}}(\theta)\ket{0}_{\mathrm{L}}
\otimes
\ket{0}_{\mathrm{anc}}
\right),
\label{eq:encoded-state-overview}
\end{equation}
where \(\ket{0}_L\) denotes the logical input and \(\ket{0}_{\mathrm{anc}}\) denotes all physical ancillas. Thus the local encoders define the truncated subspace, while the logical unitary searches for the low-energy states inside this subspace. Fig.~\ref{fig:encoded-circuit-overview} illustrates this construction for two coupled blocks.

The same construction can also be used recursively. In this case, the logical circuit $U_{\mathrm{L}}(\theta)$ will be trained within the \(k\)-dimensional retained logical subspace of the merged system. After this training, the product \(U_{\mathrm{enc}}U_{\mathrm{L}}(\theta^\star)\) is the encoder of the larger block, which maps the logical basis of the larger block into the physical low-energy subspace of the merged system.  For example, two trained $L=5$ encoders can be combined with a trained logical unitary to define a $L=10$ encoder, and two such $L=10$ encoders can then be combined in the same way to construct a $L=20$ encoder.

If \(W\) were known exactly, the projected effective Hamiltonian would be
\begin{equation}
H_{\mathrm{eff}}=W^\dagger H W.
\label{eq:logical-hamiltonian}
\end{equation}
The circuit formulation avoids explicitly constructing \(H_{\mathrm{eff}}\). Instead, the energy is estimated by preparing the physical encoded state in Eq.~\eqref{eq:encoded-state-overview} and measuring the Pauli terms of the full Hamiltonian. The following two subsections describe the two stages of the algorithm in detail.

\subsection{Ky Fan training of the local encoder}
\label{subsec:ky-fan-training}

For each block, the encoder is trained as a subspace mapping rather than as a single state-preparation circuit. Let \(H_A\) be the block Hamiltonian and let \(U_A(\eta)\) denote a parameterized encoder, where \(\eta\) is the set of variational parameters. Define the input projector
\begin{equation}
P_{\mathrm{in}}^{(A)}=
\sum_{\alpha=0}^{k-1}
\left(
\ket{\alpha}\bra{\alpha}_{\mathrm{L}}\otimes
\ket{0}\bra{0}^{\otimes(n_A-m_A)}
\right).
\label{eq:input-projector}
\end{equation}
The training cost is the Ky Fan subspace objective \cite{Fan1949Weyl,Wang2021VQSVD}
\begin{equation}
\begin{aligned}
\mathcal{C}_A(\eta)
&=
\operatorname{Tr}
\left[
P_{\mathrm{in}}^{(A)}
U_A^\dagger(\eta)
H_A
U_A(\eta)
P_{\mathrm{in}}^{(A)}
\right]
\\
&=
\sum_{\alpha=0}^{k-1}
\bra{\psi_\alpha(\eta)}
H_A
\ket{\psi_\alpha(\eta)},
\end{aligned}
\label{eq:ky-fan-cost}
\end{equation}
where
\begin{equation}
\ket{\psi_\alpha(\eta)}
=
U_A(\eta)
\left(
\ket{\alpha}_{\mathrm{L}}
\otimes
\ket{0}^{\otimes(n_A-m_A)}
\right).
\label{eq:block-training-states}
\end{equation}
Minimizing Eq.~\eqref{eq:ky-fan-cost} trains the span of the first \(k\) encoded states toward the $k$-dimensional low-energy invariant subspace of the block Hamiltonian. In a circuit implementation, each term in the sum is estimated by preparing \(\ket{\psi_\alpha(\eta)}\) and measuring the Pauli terms in the decomposition of \(H_A\).

After optimizing \(\eta\), the trained encoder \(U_A(\eta_A^\star)\) is fixed. Repeating the same training for each module gives the global encoder
\begin{equation}
U_{\mathrm{enc}}
=
\bigotimes_b U_b(\eta_b^\star),
\end{equation}
and the following optimization acts only on the logical qubits.

\subsection{Encoded variational eigensolver}
\label{subsec:encoded-vqd}
After the local encoders have been trained and fixed, the coupled-system problem is solved within the encoded tensor-product space. Let \(U_{\mathrm{L}}(\theta)\) be a parameterized logical unitary acting on the logical register of the coupled system. The encoded ground state is
\begin{equation}
\ket{\Psi(\theta)}
=
U_{\mathrm{enc}}
\left(
U_{\mathrm{L}}(\theta)\ket{0}_{\mathrm{L}}
\otimes
\ket{0}_{\mathrm{anc}}
\right),
\label{eq:encoded-state}
\end{equation}
Here \(U_{\mathrm{enc}}\) is fixed, while \(\theta\) is optimized. For a Pauli decomposition
\begin{equation}
H=\sum_{\ell=1}^{M}h_\ell P_\ell ,
\end{equation}
the energy estimator is
\begin{equation}
E(\theta)
=
\bra{\Psi(\theta)}H\ket{\Psi(\theta)}
=
\sum_{\ell=1}^{M}
h_\ell
\braket{P_\ell}_\theta .
\label{eq:energy-estimator}
\end{equation}
Each \(\braket{P_\ell}_\theta\) is estimated by preparing \(\ket{\Psi(\theta)}\), rotating the measured qubits to the appropriate Pauli basis, and measuring in the computational basis. Minimizing \(E(\theta)\) gives the encoded approximation to the ground state.

To obtain several low-energy states sequentially, we use an encoded variational quantum deflation objective \cite{Higgott2019VQD, nakanishi2019subspace}. For the \(i\)th state,
\begin{equation}
\mathcal{J}_i(\theta)
=
E(\theta)
+
\sum_{j=0}^{i-1}
\beta_{ij}F_{ij}(\theta),
\qquad
F_{ij}(\theta)
=
\left|
\braket{\Psi(\theta)|\Psi(\theta_j^\star)}
\right|^2,
\label{eq:vqd-objective}
\end{equation}
where \(\beta_{ij}>0\) are predefined penalty weights and \(\theta_j^\star\) are the parameters of previously optimized states. Because the same fixed encoder appears in every state and all ancillas are initialized in the same state, the physical overlap reduces to a logical overlap:
\begin{equation}
\braket{\Psi(\theta)|\Psi(\theta_j^\star)}
=
{}_{\mathrm{L}}\bra{0}
U_{\mathrm{L}}^\dagger(\theta)
U_{\mathrm{L}}(\theta_j^\star)
\ket{0}_{\mathrm{L}} .
\label{eq:logical-overlap}
\end{equation}
Therefore the overlap penalty can be measured on the logical register alone, without a SWAP test and without preparing two copies of the physical state. One only needs to run the single-register circuit
\[
\ket{0}_{\mathrm{L}}
\xrightarrow{\;U_{\mathrm{L}}(\theta_j^\star)\;}
\xrightarrow{\;U_{\mathrm{L}}^\dagger(\theta)\;}
\text{measure in computational basis},
\]
and record the all-zero probability
\begin{equation}
F_{ij}(\theta)
=
p_{0\cdots 0}(\theta,\theta_j^\star).
\label{eq:logical-overlap-probability}
\end{equation}
This simplification follows from the separation between the fixed block encoders and the trainable logical circuit. The complete output of the encoded eigensolver is therefore a family of state-preparation circuits
\begin{equation}
U_{\mathrm{enc}}U_{\mathrm{L}}(\theta_i^\star),
\qquad
i=0,1,\ldots,
\end{equation}
which approximate the corresponding low-energy eigenstates of the coupled Hamiltonian within the encoded subspace.

The encoded variational eigensolver described above obtains low-energy states without explicitly assembling the projected Hamiltonian. Such a method can also assist the classical modular algorithm by providing the matrix elements of \(H_{\mathrm{eff}}^{(k)}\) through circuit measurements. Once the block encoders have been trained and fixed, the projected matrix can be measured on a quantum processor and subsequently diagonalized on a classical computer.

For two blocks, we have $W=W_A\otimes W_B$ and $H_{\mathrm{eff}}^{(k)} =W^\dagger H W $. Define the logical product basis
\begin{equation}
\ket{\alpha\beta}_{\mathrm L}
=
\ket{\alpha}_{\mathrm L,A}
\otimes
\ket{\beta}_{\mathrm L,B}.
\end{equation}
If the physical Hamiltonian has the Pauli decomposition
\begin{equation}
H=\sum_{\ell}h_\ell P_\ell,
\end{equation}
its matrix elements are
\begin{equation}
\left[
H_{\mathrm{eff}}^{(k)}
\right]_{\alpha\beta,\alpha'\beta'}
=
\sum_{\ell}h_\ell
\bra{\alpha\beta}_{\mathrm L}
W^\dagger P_\ell W
\ket{\alpha'\beta'}_{\mathrm L}.
\label{eq:measured-effective-matrix}
\end{equation}

A general method for evaluating these matrix elements is the Hadamard test \cite{NielsenChuang2010}. Let \(V_{\alpha\beta}\) be a state-preparation unitary satisfying
\begin{equation}
V_{\alpha\beta}\ket{0}
=
W\ket{\alpha\beta}_{\mathrm L}.
\end{equation}
For each Pauli term \(P_\ell\), one has
\begin{equation}
\bra{0}
V_{\alpha\beta}^\dagger
P_\ell
V_{\alpha'\beta'}
\ket{0}
=
\bra{\alpha\beta}_{\mathrm L}
W^\dagger P_\ell W
\ket{\alpha'\beta'}_{\mathrm L}.
\label{eq:hadamard-transition-element}
\end{equation}
A Hadamard test applied to the unitary
\(V_{\alpha\beta}^\dagger P_\ell V_{\alpha'\beta'}\)
therefore gives the real part of this transition matrix element, and the phase-shifted version of the same test gives its imaginary part. Thus, we obtain the matrix elements of $H_{\mathrm{eff}}^{(k)}$. This procedure is general and does not require any special structure of the encoded basis, although it requires controlled implementations of the relevant state-preparation and Pauli operations.

In the present construction, a simpler procedure is available because the logical computational-basis states are orthonormal and \(W\) is an isometry. Consider two logical labels
\begin{equation}
a=(\alpha,\beta),
\qquad
b=(\alpha',\beta').
\end{equation}
We first prepare the following superposition states on the logical register
\begin{equation}
\ket{\chi_{\pm}}_{\mathrm L}
=
\frac{
\ket{a}_{\mathrm L}
\pm
\ket{b}_{\mathrm L}
}{\sqrt{2}},
\qquad
\ket{\chi_{\pm i}}_{\mathrm L}
=
\frac{
\ket{a}_{\mathrm L}
\pm
i\ket{b}_{\mathrm L}
}{\sqrt{2}} .
\label{eq:logical-superposition-states}
\end{equation}
The fixed encoder is then applied to obtain
\begin{equation}
\ket{\Psi_{\pm}}
=
W\ket{\chi_{\pm}}_{\mathrm L},
\qquad
\ket{\Psi_{\pm i}}
=
W\ket{\chi_{\pm i}}_{\mathrm L}.
\label{eq:encoded-superposition-states}
\end{equation}
For each Pauli term, define
\begin{equation}
M_{ab}^{(\ell)}
=
\bra{a}_{\mathrm L}
W^\dagger P_\ell W
\ket{b}_{\mathrm L}.
\end{equation}
Its real and imaginary parts can be reconstructed from ordinary expectation-value measurements,
\begin{equation}
\operatorname{Re}M_{ab}^{(\ell)}
=
\frac{
\bra{\Psi_+}P_\ell\ket{\Psi_+}
-
\bra{\Psi_-}P_\ell\ket{\Psi_-}
}{2},
\label{eq:real-offdiag-measurement}
\end{equation}
and
\begin{equation}
\operatorname{Im}M_{ab}^{(\ell)}
=
\frac{
\bra{\Psi_{-i}}P_\ell\ket{\Psi_{-i}}
-
\bra{\Psi_{+i}}P_\ell\ket{\Psi_{+i}}
}{2}.
\label{eq:imag-offdiag-measurement}
\end{equation}
The diagonal elements are obtained by simply preparing
\(W\ket{\alpha\beta}_{\mathrm L}\)
and measuring the expectation values of the Pauli terms \(P_\ell\). Combining these measurements with the coefficients \(h_\ell\) reconstructs the full projected Hamiltonian in Eq.~\eqref{eq:measured-effective-matrix} \cite{McClean2017QSE,Colless2018MolecularSpectra}.

\begin{figure*}
\centering
\includegraphics[width=0.95\textwidth]{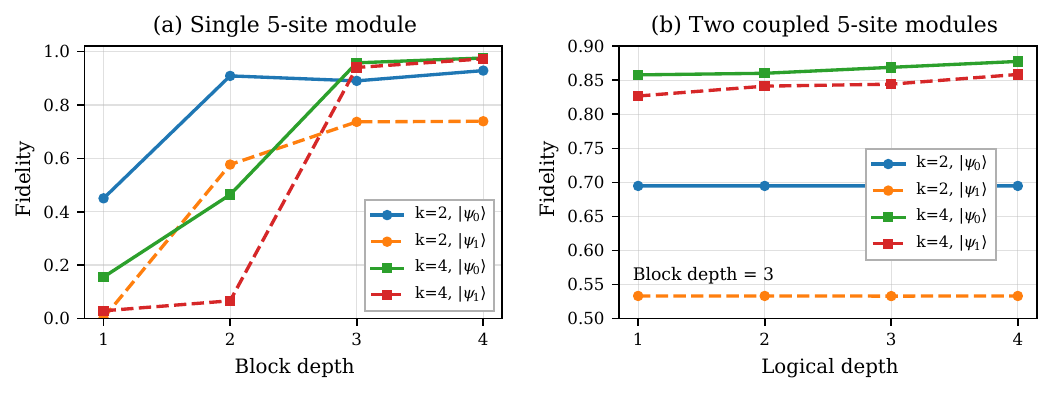}
\caption{
Numerical training results for the encoded-subspace construction at \(J=1.0\) and \(h=1.0\). (a) shows the fidelities for a single $L=5$ module for subspace dimensions \(k=2\) and \(k=4\). (b) shows the fidelities for the $L=10$ system obtained by coupling two $L=5$ modules, with the block depth fixed to \(3\). Here, \( |\psi_0\rangle \) and \( |\psi_1\rangle \) denote the target ground state and first excited state, respectively. We can see that the \(k=4\) encoded subspace yields substantially better performance than the \(k=2\) encoded subspace for both the ground-state and first-excited-state.
}
\label{fig:training-results}
\end{figure*}

This measurement-assisted approach connects the quantum implementation to the classical projected eigensolver. The quantum processor prepares the encoded states and supplies the projected matrix elements, while the resulting \(k^2\times k^2\) Hamiltonian can be diagonalized classically. Such circuit representation provides a substantial storage advantage. In the classical implementation, the retained block states are stored as explicit amplitude vectors in the physical Hilbert space, and this storage cost grows with the block size. While in the circuit implementation, the same low-energy subspace is represented by the parameters of the trained encoder, whose number is much smaller than the number of amplitudes of the state vectors. Once \(U_A\) and \(U_B\) have been trained, the corresponding basis states are generated by applying these circuits to logical basis states with fixed ancillas, rather than by loading or storing all state amplitudes. Thus the representation of the subspace is shifted from explicit state storage to compact circuit-parameter storage.

\subsection{Qiskit numerical training results}
\label{subsec:qiskit-results}

We finally use the transverse-field Ising model to verify our hardware implementation method. Here we use Qiskit to build and train the corresponding quantum circuits~\cite{JavadiAbhari2024Qiskit,QiskitCommunity2019}. In this procedure, we first train an \(L=5\) block encoder to represent the lowest \(k\) block eigenstates. Using this trained block encoder as the elementary module, we then construct larger encoded systems and calculate the low-energy spectra for \(L=10,15,20,25,30\).

In this demonstration, the block encoder and the logical circuit are composed of hardware-efficient layers \cite{Kandala2017HardwareEfficient}. For an \(n\)-qubit register, one layer consists of a product of single-qubit rotations followed by a fixed nearest-neighbor entangling pattern,
\begin{equation}
U_{\mathrm{lay}}(\vartheta)
=
U_{\mathrm{ent}}
\left[
\bigotimes_{q=1}^{n}
R_z(\vartheta_{q,2})R_y(\vartheta_{q,1})
\right],
\end{equation}
where \(U_{\mathrm{ent}}\) is a brick-wall pattern of nearest-neighbor CNOT gates (as illustrated in Fig.~\ref{fig:trained-circuits-k2-k4}). The variational parameters \(\vartheta\) are independently optimized for each layer. A depth-\(D\) ansatz is then defined as
\begin{equation}
U^{(D)}(\vartheta)
=
U_{\mathrm{lay}}(\vartheta_D)
U_{\mathrm{lay}}(\vartheta_{D-1})
\cdots
U_{\mathrm{lay}}(\vartheta_1).
\end{equation}
Throughout this section, ``depth'' denotes the number of repeated variational layers in the above ansatz. The block depth \(D_{\mathrm{b}}\) refers to the number of variational layers used in the five-qubit block encoder \(U_{A}\), and the logical depth \(D_{\mathrm{L}}\) denotes the number of variational layers in this logical circuit \(U_{\mathrm{L}}\), which acts on the logical qubits of all blocks. 

The training results are shown in Fig.~\ref{fig:training-results}. Here, we set \(J=h=1\) for the circuit-level demonstrations because the case at the critical point provides a relatively demanding benchmark for low-energy-state preparation. For $k=2$, increasing the block depth rapidly improves the block ground-state fidelity from roughly $0.45$ to above $0.90$, while the first-excited-state fidelity increases from nearly zero to about $0.74$. Note that at small block depths, the smaller encoded subspace with \(k=2\) can perform better than the larger one with \(k=4\). The advantage of a larger $k$ becomes visible only when the block depth is sufficiently increased. In principle, increasing \(k\) enlarges the accessible low-energy subspace and raises the best achievable fidelity. However, a larger \(k\) also requires a more expressive circuit ansatz. Therefore, to obtain the fidelity improvement expected from a larger retained subspace, the block depth must be increased accordingly. In the present calculation, for \(k=4\), both displayed fidelities exceed \(0.94\) once the block depth reaches \(3\), indicating that this depth is sufficient to learn a reliable local low-energy encoded subspace for the chosen ansatz and parameters.

After coupling two trained $L=5$ modules, the block encoders are held fixed and only the logical circuit is optimized. With block depth fixed to $3$, the $k=2$ construction gives an almost depth-independent $L=10$ ground-state fidelity near $0.695$, indicating that two retained states per module are a little restrictive for the coupled system. The $k=4$ construction gives a higher fidelity, increasing to $0.878$ for ground state and $0.859$ for first-excited state. This behavior confirms the classical projection results in the previous sections, showing that on hardware, we can indeed implement this truncation method and obtain good low-energy states with only modest circuit depth. 

\begin{figure}
\centering
\includegraphics[width=\columnwidth]{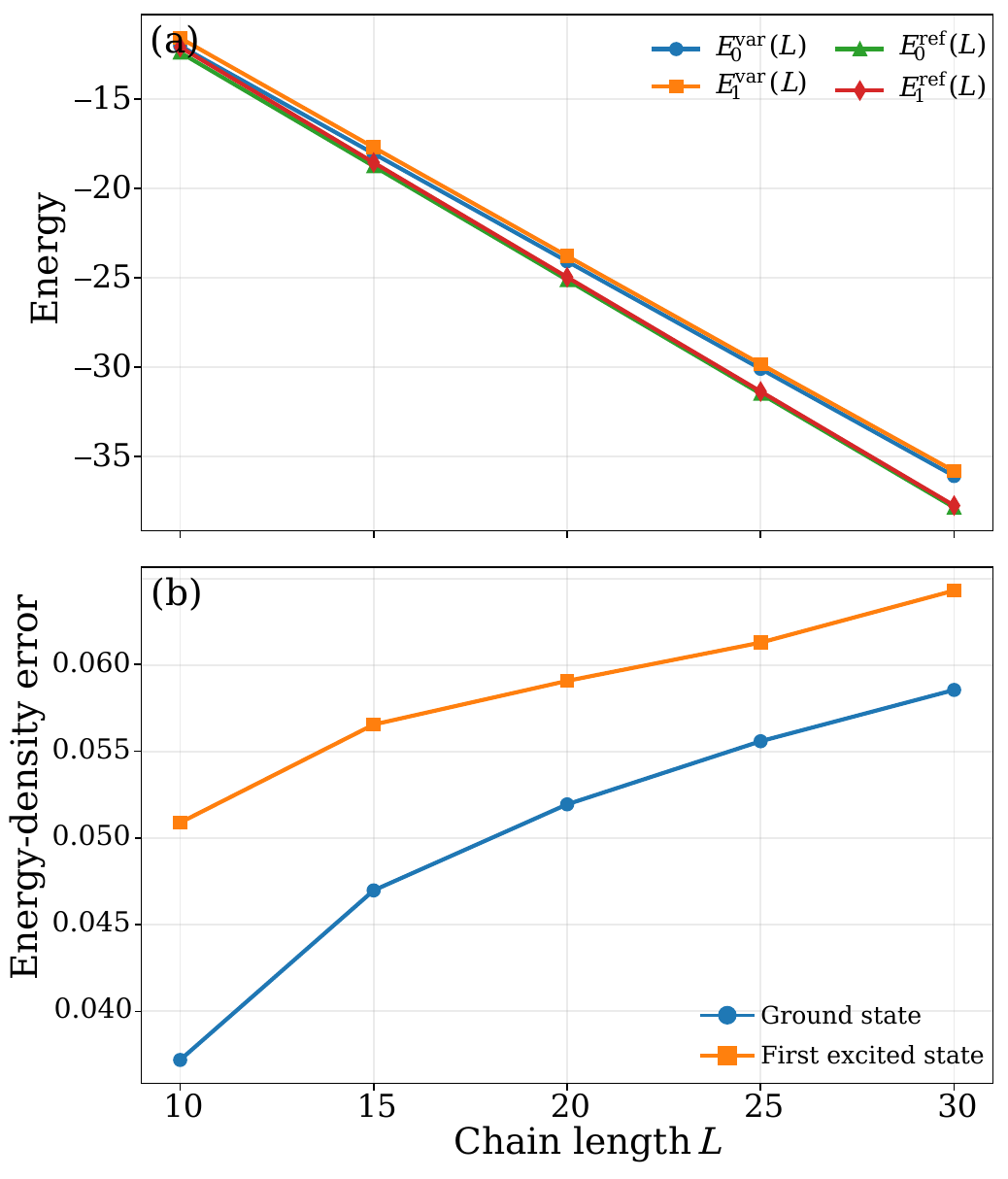}
\caption{Scale-up test based on a multi-module encoded construction with fixed truncation level \(k=2\) at \(J=1.0\) and \(h=1.0\). A single trained $L=5$ encoder is reused for each module, and the number of modules is increased to produce chains of length \(N=10,15,20,25,30\). (a) shows the variational energies \(E_0^{\mathrm{var}}(N)\) and \(E_1^{\mathrm{var}}(N)\) for the ground state and the first excited state, together with the corresponding reference energies \(E_0^{\mathrm{ref}}(N)\) and \(E_1^{\mathrm{ref}}(N)\) obtained from the exact free-fermion solution. (b) shows the corresponding energy-density errors \((E_0^{\mathrm{var}}-E_0^{\mathrm{ref}})/N\) and \((E_1^{\mathrm{var}}-E_1^{\mathrm{ref}})/N\). The variational energies track the reference energies well as the chain length increases, while the energy-density errors remain at the level of \(10^{-2}\) and grow only gradually with \(N\). This indicates that the encoded multi-module construction remains effective for larger systems.
}
\label{fig:multimodule-scaleup}
\end{figure}

As a complementary scale-up test, we have also considered a multi-module encoded construction with fixed truncation level \(k=2\). The purpose is to test whether our method is still effective when a logical variational circuit captures correlations among an increasing number of modules. In this construction, an \(L=5M\) chain is assembled from \(M\) elementary \(L=5\) modules. The same trained \(L=5\) encoder is applied to every module, so each module is represented by one logical qubit, and \(M\) is increased to obtain \(L=10,15,20,25,30\) chains. We then optimize the logical circuit over all \(M\) logical qubits for the ground and first excited states, and compare the resulting energies with the corresponding reference values.

The results are shown in Fig.~\ref{fig:multimodule-scaleup}. Figure (a) shows that both the variational ground-state energy and the variational first-excited-state energy follow the corresponding reference energies closely as the chain length increases. Figure (b) further shows that the energy-density errors remain well controlled throughout the tested range. Although the total energy error accumulates with system size, the error per site remains at the level of \(10^{-2}\) and grows only gradually. This behavior provides clear evidence that the encoded-subspace circuit construction can be extended beyond the $L=10$ setting and still yields meaningful low-energy approximations for larger chains.

\subsection{Hardware execution on IBM quantum processors}

To physically prepare the encoded low-energy states and confirm the performance of our method on real quantum hardware, we executed the trained state-preparation circuits on IBM quantum processors. The parameters and the target Hamiltonian are the same as in Section~\ref{subsec:qiskit-results}. We considered both the ground state \(\ket{\psi_0}\) and the first excited state \(\ket{\psi_1}\), and compared the hardware performance for different subspace dimensions \(k\) and local block depths. 

The resulting ground-state and first-excited-state preparation circuits were transpiled and executed on the IBM backend \texttt{ibm\_pittsburgh}. All hardware results were obtained directly from the measured outcomes without error-mitigation procedure.
This avoids the additional sampling and post-processing costs associated with error mitigation and provides a direct assessment of the encoded-state-preparation procedure under the native hardware noise. Since the hardware measurement was performed in the computational basis, it does not give us access to the full quantum-state fidelity. Instead, we can only obtain the fidelity between bit-string probability distributions. Therefore, we consider the measured distribution fidelity~\cite{bhattacharyya1943measure}
\begin{equation}
F_Z
=
\left(
\sum_z
\sqrt{
p_{\mathrm{ibm}}(z)\,
p_{\mathrm{ref}}(z)
}
\right)^2 .
\end{equation}
Here, $p_{\mathrm{ref}}(z)=|\langle z|\Phi\rangle|^2$ is the exact computational-basis distribution for a reference state, and \(p_{\mathrm{ibm}}(z)\) is the empirical distribution obtained from IBM hardware. This quantity equals one when the measured computational-basis distribution exactly matches the reference distribution. It should be distinguished from the full state fidelity, since relative phase information and off-diagonal coherences are not reconstructed from computational-basis measurements alone.

\begin{figure}
\centering
\includegraphics[width=\columnwidth]{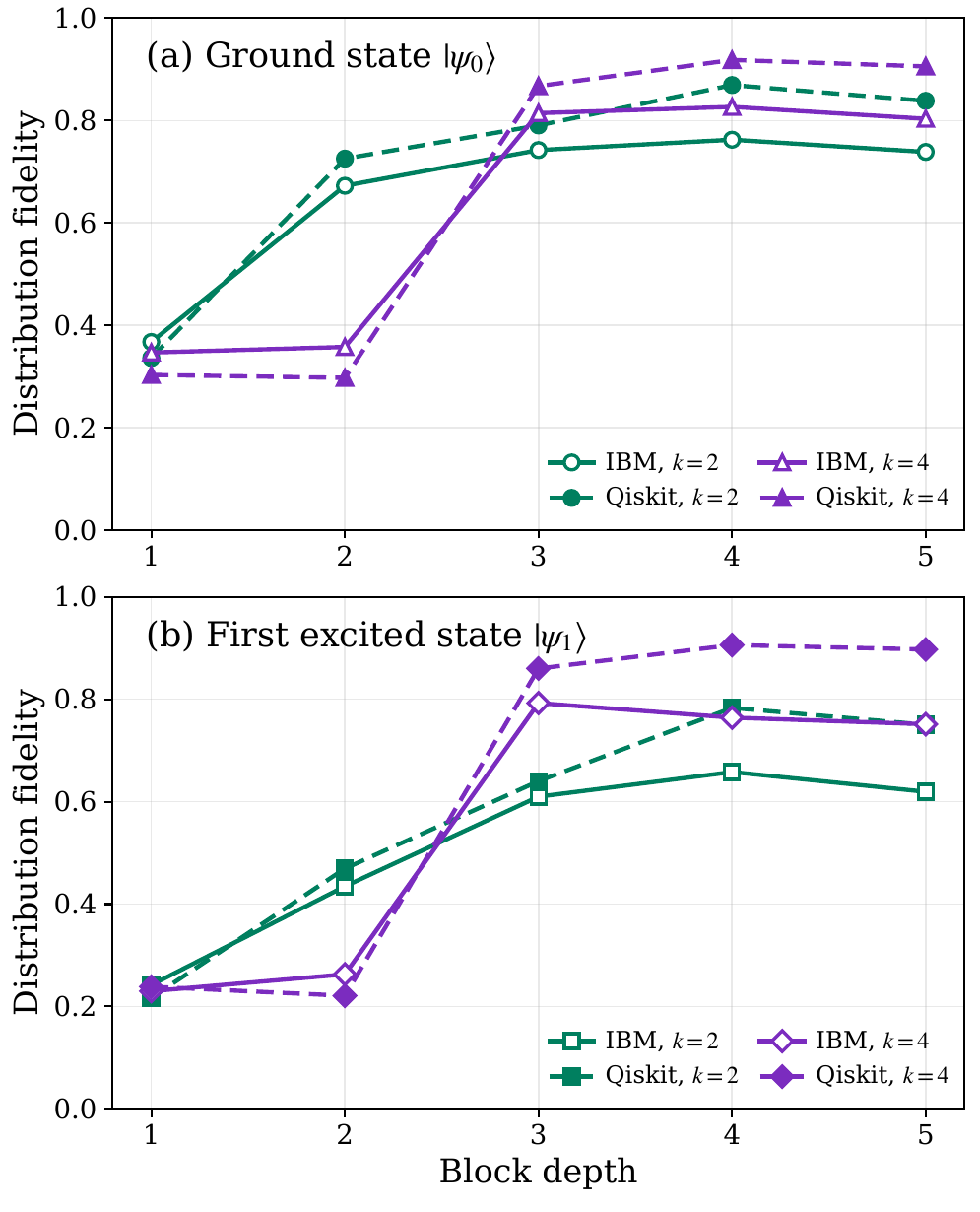}
\caption{
IBM hardware and noiseless Qiskit results for the $L=10$ transverse-field Ising model at \(J=h=1.0\), with the logical depth fixed to \(1\). (a) shows the computational-basis distribution fidelity \(F_Z\) for the ground state \(\ket{\psi_0}\), and (b) shows the corresponding fidelity for the first excited state \(\ket{\psi_1}\). Solid curves with open markers denote IBM hardware results, while dashed curves with filled markers denote noiseless Qiskit results for the same state-preparation circuits. The Qiskit results generally provide higher fidelities at intermediate and large depths, while the IBM results follow the same overall depth dependence.
}
\label{fig:qiskit-ibm-fidelity-comparison}
\end{figure}

For each \(k\), we evaluated the hardware distribution fidelity for both the ground-state circuit and the first-excited-state circuit. The distribution fidelity with respect to the exact eigenstates are shown in Fig.~\ref{fig:qiskit-ibm-fidelity-comparison}.

The data in Fig.~\ref{fig:qiskit-ibm-fidelity-comparison} show that increasing the block depth can improve the fidelity when the circuit is not too deep. For \(k=2\) case, the fidelities increase steadily as the block depth is increased from \(1\) to \(4\), with the ground-state fidelity rising to approximately \(0.76\) and the first-excited-state fidelity rising to \(0.66\). At depth \(5\), both fidelities decrease slightly. For \(k=4\), the fidelities attain their best values when the block depth reaches \(3\), approximately \(0.83\) for the ground state and \(0.76\) for the first excited state, exceeding the corresponding \(k=2\) values. These results confirm that reasonably high fidelity can already be achieved with a relatively small value of $k$.

The noiseless Qiskit results provide an ideal fidelity for the trained circuits. In this ideal setting, increasing the block depth generally improves the circuit performance. However, the IBM hardware results show a different behavior. In the present data, increasing the block depth can reduce the measured distribution fidelity, especially when the depth becomes too large. This discrepancy is mainly caused by hardware noise. A deeper encoder requires more physical gates after transpilation, and the accumulated gate errors, decoherence, and readout errors can outweigh the improvement obtained from the more expressive variational ansatz~\cite{sim2019expressibility}. Therefore, on present noisy quantum processors, the optimal circuit depth is determined by a balance between ansatz expressibility and hardware noise.

\begin{figure}
\centering
\includegraphics[width=\columnwidth]{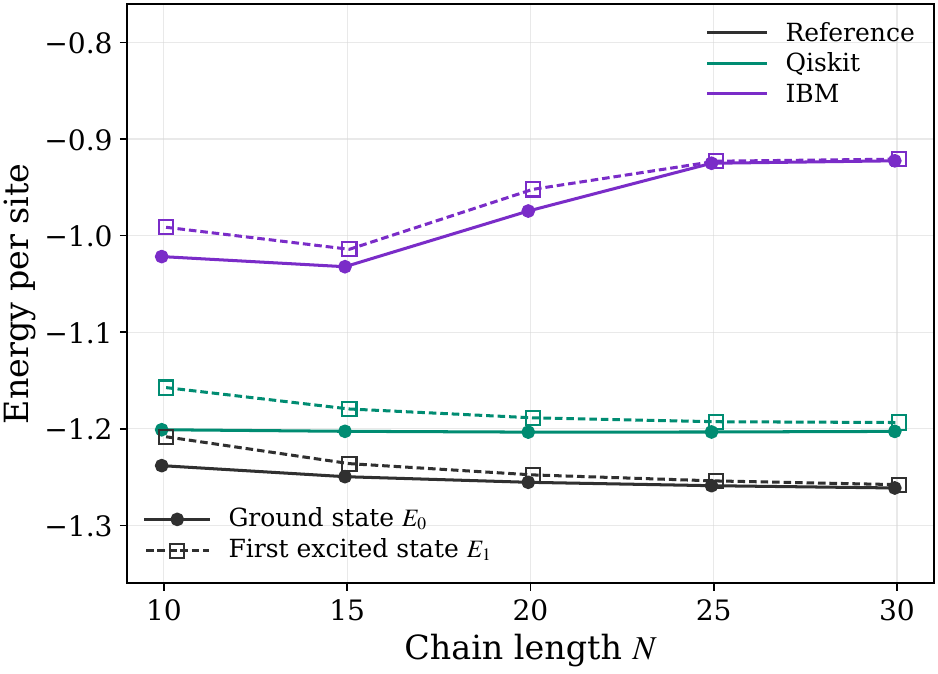}
\caption{
Ground-state and first-excited-state energies per site for transverse-field Ising chains of different lengths at $J=h=1.0$, calculated using the modular construction with a fixed truncation level $k=2$. We compare the variational estimates obtained from noiseless Qiskit simulations and IBM hardware measurements with the corresponding reference values. The noiseless Qiskit results remain close to the reference values throughout the tested range of chain lengths, whereas the IBM hardware results show larger deviations.}
\label{fig:multimodule-qiskit-ibm-energy}
\end{figure}

The energy estimation results for larger systems are shown in Fig.~\ref{fig:multimodule-qiskit-ibm-energy}. In the noiseless simulation, both \(E_0^{\mathrm{Qiskit}}\) and \(E_1^{\mathrm{Qiskit}}\) follow the reference energies closely as the chain length increases, while in the IBM results, the absolute energies are significantly higher than the reference values. The reduced accuracy of the IBM energy estimates mainly comes from two sources. First, hardware noise during state preparation, including gate errors and decoherence, causes the prepared state to deviate from the target encoded low-energy state. Second, the measurement process introduces additional errors, including readout errors and errors from the basis rotations required to measure different Pauli terms~\cite{bravyi2021mitigating}. These two effects lead to inaccurate estimates of the individual Pauli expectation values. Since the final energy is obtained as a weighted sum of many such Pauli terms, the errors from different terms accumulate in the total energy estimate. This accumulation becomes more severe for larger \(L\), where the circuits contain more qubits and more compiled gates, and the Hamiltonian contains more measured contributions. Therefore, the IBM results should be viewed mainly as a hardware-level demonstration that the encoded circuits can be executed for larger multi-module systems, while the noiseless Qiskit results better represent the intrinsic accuracy of the variational encoded-subspace construction.

These results demonstrate that the encoded-subspace construction can be implemented as an actual state-preparation protocol on IBM quantum processors. Despite the presence of hardware noise and the additional compilation overhead, the trained circuits are able to prepare the low-energy states of the transverse-field Ising model with appreciable fidelity.  This provides clear evidence that the proposed circuit-level implementation can produce meaningful low-energy states on present-day quantum processors, suggesting that our variational method can be a hardware-compatible route to approximating many-body low-energy states. 

\section{Conclusion}
In this work, we investigate a recursive, physically intuitive module-coupling algorithm for computing low-energy many-body eigenstates. The simple idea is to iteratively construct low-dimensional subspaces from the low-energy eigenstates of smaller modules. When the inter-module coupling does not strongly deform this low-energy product space, the low-lying eigenstates of the full system can be well represented in the subspace with high accuracy. The resulting projected Hamiltonian is therefore much easier to diagonalize.

We carefully benchmark our recursive multi-state eigensolver on transverse-field Ising chains using classical numerical calculations. For the $L=10$ system, it confirms that a relatively small retained subspace is sufficient to reproduce the low-energy spectrum with high fidelity. By testing different coupling geometries we showed that our physics-informed approach remains robust under changes in the coupling structure. We then apply this construction recursively to attack longer chains. The low-energy spectra, ground-state fidelities, excitation energy gaps, cost analysis and energy-error scaling all indicate that the retained module subspaces provide a compact and physically meaningful representation of the low-energy subspace.  Results from this part can be already highly useful for reliable and computationally inexpensive estimates of the gap between the ground state and excited states of a many-body quantum system.

Owing to its recursive and modular structure, our approach can be mapped naturally onto current gate-based quantum hardware. This compatibility makes the direct preparation of low-energy quantum states feasible and motivates the development of a quantum-circuit realization of the modular construction. In this formulation, trained block encoders represent the retained local subspaces, while logical variational circuits couple these encoded subspaces and search for the low-energy states of successively larger blocks. The resulting encoders can be reused recursively as elementary circuit modules at the next level, thereby extending the classical modular projected eigensolver into a hierarchical quantum state-preparation protocol. Our noiseless simulations show that this encoded construction can reproduce low-energy states with modest circuit depth, while results obtained on IBM quantum hardware demonstrate its implementability on present-day quantum processors and yield appreciable computational-basis distribution fidelities despite hardware noise.

The present work also suggests several directions for future study. First, this work offers an alternative way to construct low-dimensional subspaces for many-body eigensolvers. Instead of relying on a generic variational ansatz in the full Hilbert space, the subspace is built from physically motivated module eigenstates and then recursively extended to larger systems. This is different from methods such as DMRG, where the effective variational representation is obtained through repeated tensor optimizations and sweeps. It may motivate the development of other subspace-construction strategies, including adaptive choices of modules, optimized selections of retained states, or hybrid subspaces that combine local eigenstates with other physically informed basis states. More broadly, our results suggest that low-energy subspaces themselves can serve as reusable computational objects. Local spectral information can be recursively assembled into larger low-energy manifolds, used directly for efficient classical eigensolving, and encoded as modular quantum circuits for physical state preparation. This classical-to-quantum continuity is a central advantage of the recursive module-coupling framework and provides a natural route for combining approximate eigensolving with subsequent quantum refinement algorithms.

Second, in many quantum algorithms and physical applications, estimating the energy alone is insufficient. One must also prepare the corresponding low-energy state as an actual quantum state for subsequent measurements, dynamical simulations, or further quantum-information processing. Our noisy hardware implementation demonstrates that the encoded-subspace construction can serve not only as a classical approximation scheme, but also as a practical route for preparing low-energy many-body states on quantum devices. To achieve higher fidelity, error-mitigation techniques may be incorporated in future quantum implementations. This feature could be particularly useful for studies that rely on high-quality initial states, where an accurate and compact state-preparation protocol is essential. 

As future work we should leverage on the compatibility between this work and several existing algorithms to treat problems at even larger scales. In particular, accurately obtaining low-lying eigenstates of each module to be coupled is essential to scale up our algorithm.  When the starting point is already a relatively large module for which exact diagonalization is not feasible, DMRG may be first used to obtain the low-energy subspace before applying our algorithm to significantly increase the system size further.  Likewise,  because our approach is computationally simple, the resulting low-energy eigenstates can serve as initial states for other algorithms, such as DMRG or imaginary-time evolution.

\begin{acknowledgments}

  J.G. acknowledges support by the National Research Foundation, Singapore, through the National Quantum Office, hosted in A*STAR, under its Centre for Quantum Technologies Funding Initiative (S24Q2d0009). C.H. acknowledges support by the Singapore Ministry of Education (MOE) through the Academic Research Fund Tier-II Grant (MOE-T2EP50224-0007).
We acknowledge the use of IBM Quantum services for this work. The views expressed are those of the authors, and do not reflect the official policy or position of IBM or the IBM Quantum team.
\end{acknowledgments}

\appendix

\section{Computational-complexity derivation}
\label{app:complexity}
In the main text, we identified the central features of the computational cost: each merge requires the diagonalization of an effective Hamiltonian in a $k^2$-dimensional product basis, while a homogeneous chain of length $L$ is constructed through only $\log_2(L/L_0)$ distinct merge levels. This appendix provides the detailed time- and memory-complexity analysis underlying these estimates, including the initial preparation of the elementary modules.

Let $d$ be the on-site Hilbert-space dimension, let $L_0$ be the elementary
module length, and let $M=L/L_0$ be the number of elementary modules.  The
Hilbert-space dimension of one elementary module is $D_0=d^{L_0}$.  Dense
diagonalization of one such module therefore requires
$\mathcal{O}(d^{3L_0})$ operations and $\mathcal{O}(d^{2L_0})$ memory.

At every recursive merge, the product basis has dimension $D=k^2$.  Forming a
dense effective Hamiltonian with a fixed number of boundary-coupling terms
requires $\mathcal{O}(k^4)$ operations and $\mathcal{O}(k^4)$ memory.  The
present implementation computes the full eigendecomposition of this matrix,
which costs $\Theta(D^3)=\Theta(k^6)$.  It also forms the projected boundary
operators explicitly.  Their dense basis transformations cost at most
$\mathcal{O}(k^5)$ and are therefore asymptotically subleading.  Consequently,
the time and working-memory costs of one merge in the implementation used here
are
\begin{equation}
 T_{\mathrm{merge}}=\Theta(k^6),
 \qquad
 S_{\mathrm{merge}}=\Theta(k^4).
 \label{eq_merge_complexity}
\end{equation}

For the homogeneous setting considered here, all blocks at a
given level are identical and the same renormalized block is reused. To obtain a chain with length $L=ML_0$, the algorithm contains $\log_2 M$ merge problems in total, thus the cost for the whole merge process will be
$\mathcal{O}\!\left(k^6\log_2 M\right)$, after the elementary modules have
been prepared. The total serial work is therefore
\begin{equation}
 T_{\mathrm{total}}
 =\mathcal{O}\!\left(d^{3L_0}+k^6\log_2 M\right).
 \label{eq_homogeneous_complexity}
\end{equation}
The working memory is
$\mathcal{O}\!\left(d^{2L_0}+k^4\right)$. Consequently, for homogeneous systems, the logarithmic dependence on $L$ allows the low-energy spectrum of long chains to be obtained rapidly when the retained dimension $k$ is modest.

\section{Dependence on coupling geometry}
\label{app:coupling-geometry}

\begin{figure}
\centering
\includegraphics[width=0.95\columnwidth]{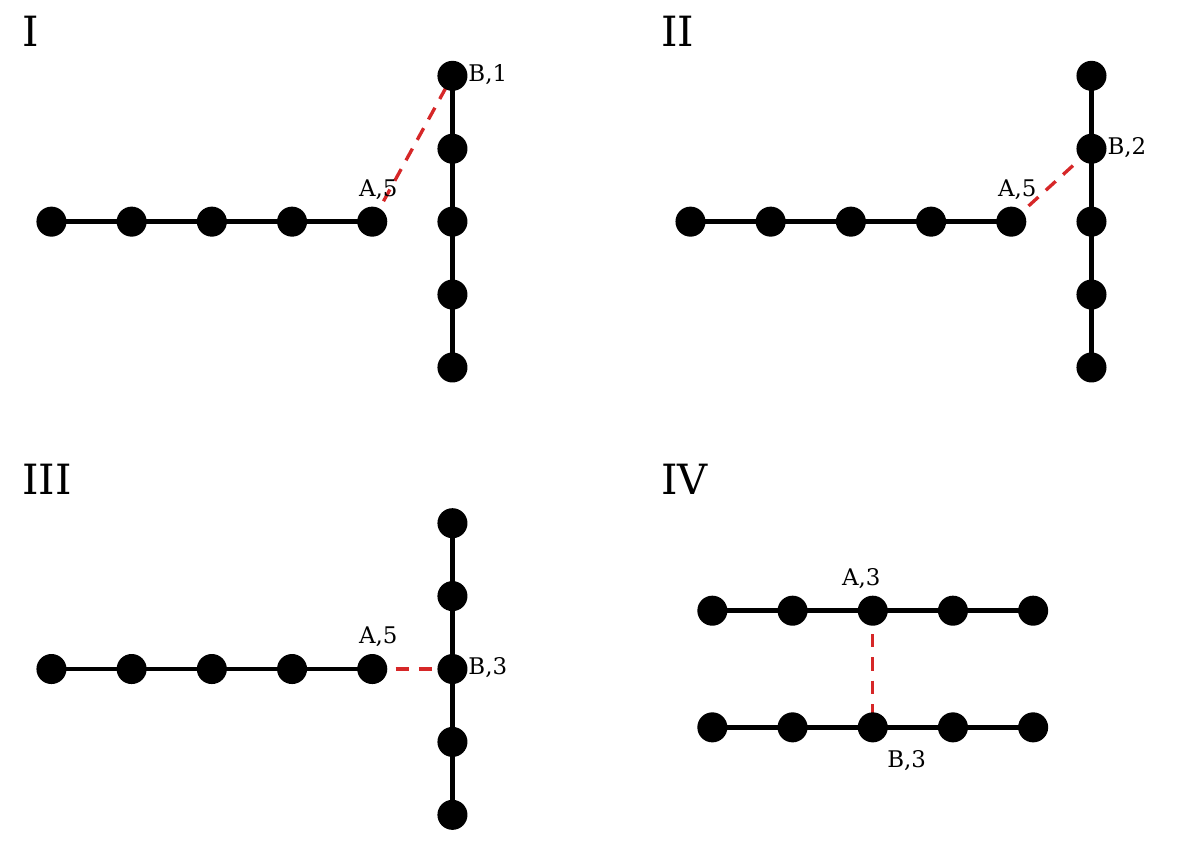}
\caption{
Schematic illustration of the four inter-module coupling geometries considered in this work. The four panels correspond to
I: \((i,j)=(5,1)\),
II: \((i,j)=(5,2)\),
III: \((i,j)=(5,3)\),
and
IV: \((i,j)=(3,3)\).}
\label{fig:coupling-geometries}
\end{figure}

\begin{figure*}
\centering
\includegraphics[width=0.95\textwidth]{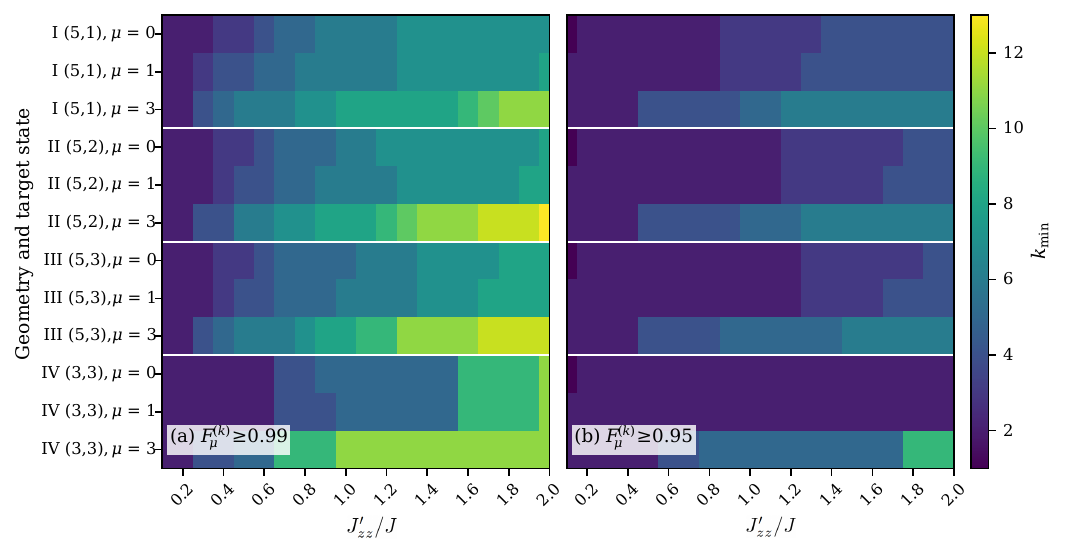}
\caption{
Minimal subspace dimension \(k_{\min}\) (per module) required to approximate the selected low-energy eigenstates for \(J=1.0\) and \(h=0.7\). (a) shows the threshold \(F_\mu^{(k)}\ge 0.99\), while (b) shows the threshold \(F_\mu^{(k)}\ge 0.95\).
The horizontal axis denotes the inter-module coupling ratio \(J_{zz}'/J\), and the vertical axis labels both the coupling geometry and the target eigenstate. Here $\mu=0,1,3$ correspond to the ground state, the first excited state, and the third excited state respectively.  
The color scale gives the smallest local subspace dimension \(k_{\min}\) required to reach the prescribed fidelity.
}
\label{fig:fidelity-coupling-position}
\end{figure*}

The main-text benchmarks employ the standard nearest-neighbor coupling across the adjacent boundaries of two modules. Since the inter-module coupling determines which block excitations are admixed into the low-energy states of the combined system, this appendix examines the robustness of the modular truncation under different coupling positions. Each module considered here is described by the same five-site TFIM Hamiltonian used in Sec.~\ref{subsec:ten-site-benchmark}
\begin{equation}
H_{A(B)}=-J\sum_{i=1}^{L-1}\sigma_i^z\sigma_{i+1}^z-h\sum_{i=1}^{L}\sigma_i^x .
\label{eq:block-hamiltonian}
\end{equation}
The coupling geometries are chosen as (see Fig.~\ref{fig:coupling-geometries}):
\begin{align}
\mathrm{I}:\quad &V=-J_{zz}'\,\sigma_{A,5}^z\otimes\sigma_{B,1}^z,\\
\mathrm{II}:\quad &V=-J_{zz}'\,\sigma_{A,5}^z\otimes\sigma_{B,2}^z,\\
\mathrm{III}:\quad &V=-J_{zz}'\,\sigma_{A,5}^z\otimes\sigma_{B,3}^z,\\
\mathrm{IV}:\quad &V=-J_{zz}'\,\sigma_{A,3}^z\otimes\sigma_{B,3}^z.
\end{align}
For each case, we determine how many eigenstates per module are required to reach a prescribed fidelity for selected low-energy states. The results are shown in Fig.~\ref{fig:fidelity-coupling-position}.

\begin{figure*}
\centering
\includegraphics[width=0.98\textwidth]{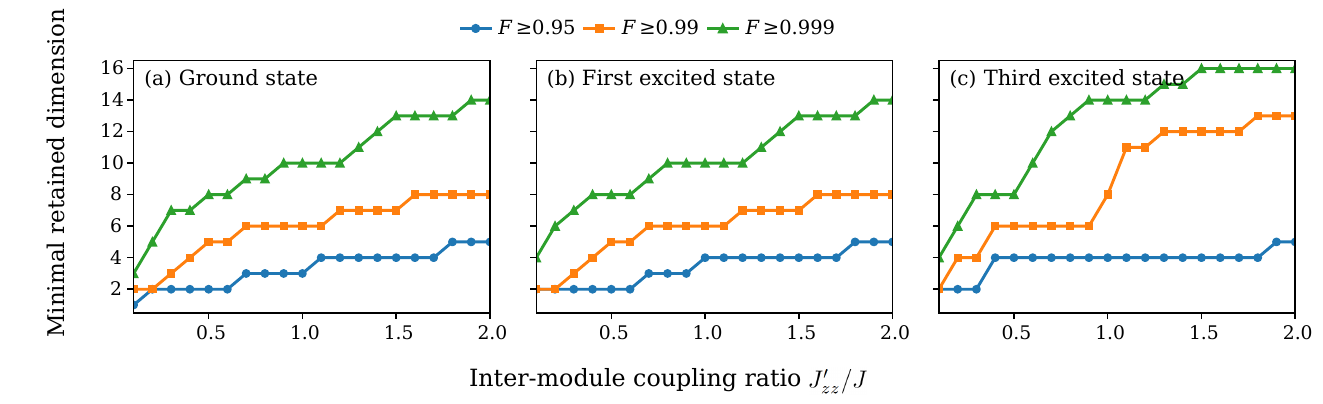}
\caption{
Minimal subspace dimension per module required to reach the prescribed fidelity thresholds for the next-nearest-coupling case, obtained at \(J=1.0\) and \(h=0.7\).
The three panels correspond to the ground state, the first excited state, and the third excited state, respectively.
The inter-module coupling is taken as Eq.~(\ref{eq:next-nearest-coupling}) with \(\gamma=0.2\).
For each target state, the plotted curves show the minimal subspace dimension required to achieve \(F\ge 0.95\), \(F\ge 0.99\), and \(F\ge 0.999\), respectively.
}
\label{fig:next-nearest-coupling-fidelity}
\end{figure*}

These results further confirm the convergence of the modular truncation. For all the coupling geometries, a subspace dimension of \(k\approx 10\) is already sufficient to achieve high fidelity (\(F^{(k)}_\mu>0.99\)). Figure~\ref{fig:fidelity-coupling-position} shows that end-to-end and near-edge couplings can be captured with smaller \(k\), whereas central couplings generally require a larger retained subspace. This trend is consistent with the intuition that a local perturbation acting near the center of a finite block can couple more efficiently to internal block excitations.

We also consider a coupling containing small additional next-nearest contributions,
\begin{equation}
V=-J_{zz}'\left(
\sigma_{A,5}^z\otimes\sigma_{B,1}^z
+\gamma\,\sigma_{A,4}^z\otimes\sigma_{B,1}^z
+\gamma\,\sigma_{A,5}^z\otimes\sigma_{B,2}^z
\right),
\label{eq:next-nearest-coupling}
\end{equation}
with $\gamma=0.2$. The corresponding fidelities are shown in Fig.~\ref{fig:next-nearest-coupling-fidelity}.

Compared with the single nearest-neighbor link, the weak next-nearest couplings lead to a reduction in fidelity at fixed $k$. Equivalently, a slightly larger local subspace is needed to reach the same accuracy. This is expected as the coupling contains longer-distance interactions and can therefore admix a broader set of block excitations into the low-energy states of the coupled system. However, the required minimal subspace dimension \(k_{\min}\) remains modest, showing that the low-energy modular basis is still compact even in the presence of these additional coupling terms.

\section{Detailed truncation-error analysis}
\label{app:error-bounds}

This appendix provides the detailed derivations underlying the truncation-error bounds summarized in Sec.~\ref{sec:error-bounds}, together with additional numerical tests of the single-merge bound and the dependence on the elementary module size.

Let $P=P_k$ project onto the retained product sector and let $Q=I-P$. With respect to the decomposition of the Hilbert space into the retained and discarded sectors, the Hamiltonian has the block form
\begin{equation}
\begin{aligned}
H&=
\begin{pmatrix}
A & B\\
B^\dagger & C
\end{pmatrix},\\
A&=PHP,
\qquad
B=PHQ,
\qquad
C=QHQ.
\end{aligned}
\label{eq_single_merge_blocks}
\end{equation}
Define $a=\lambda_{\min}(A)$, $b=\lVert B\rVert_2$, and suppose that the discarded block is separated from the projected ground energy by a positive gap
\begin{equation}
\lambda_{\min}(C)-\lambda_{\min}(A)
\geq
\gamma,
\qquad
\gamma>0.
\label{eq_single_merge_gap}
\end{equation}
For an arbitrary normalized state
$|\psi\rangle=|x\rangle+|y\rangle$, where
$|x\rangle=P|\psi\rangle$ and $|y\rangle=Q|\psi\rangle$, Eqs.~\eqref{eq_single_merge_blocks} and \eqref{eq_single_merge_gap} give
\begin{align}
\langle\psi|H|\psi\rangle
&\geq
a\lVert x\rVert^2
+
(a+\gamma)\lVert y\rVert^2
-
2b\lVert x\rVert\lVert y\rVert
\notag\\
&=
a+
\begin{pmatrix}
\lVert x\rVert & \lVert y\rVert
\end{pmatrix}
\begin{pmatrix}
0 & -b\\
-b & \gamma
\end{pmatrix}
\begin{pmatrix}
\lVert x\rVert\\
\lVert y\rVert
\end{pmatrix}.
\label{eq_single_merge_rayleigh}
\end{align}
The smaller eigenvalue of the last $2\times2$ matrix is
$(\gamma-\sqrt{\gamma^2+4b^2})/2$. Minimizing Eq.~\eqref{eq_single_merge_rayleigh} and using the variational inequality $E_0(H)\leq a$ therefore yields
\begin{align}
0\leq a-E_0(H)
&\leq
r(b,\gamma)
=
\frac{\sqrt{\gamma^2+4b^2}-\gamma}{2}
\notag\\
&=
\frac{2b^2}
{\sqrt{\gamma^2+4b^2}+\gamma}
\leq
\frac{b^2}{\gamma}.
\label{eq_single_merge_bound}
\end{align}

Since the uncoupled module Hamiltonian does not couple the retained and discarded sectors, $P(H_A\otimes I_B+I_A\otimes H_B)Q=0$,
and hence $B=PHQ=PVQ$.
Therefore, $b=\lVert B\rVert_2$ measures the strength of the mixing between the retained and discarded sectors generated by the inter-module coupling. For the bounded local coupling considered here,
$b\leq\lVert V\rVert$ is finite and independent of $L$. The quantity $\gamma$ characterizes the energy separation of the discarded sector from the low-energy subspace, namely, the difference between the lowest energies of $PHP$ and $QHQ$. Equation~\eqref{eq_single_merge_bound} therefore indicates that weaker retained--discarded mixing and a larger energy separation reduce the truncation error introduced by a single merge. Under the assumed nonzero discarded-sector separation, $b^2/\gamma$ remains finite throughout the recursive construction.

\begin{figure}[t]
\centering
\includegraphics[width=\columnwidth]{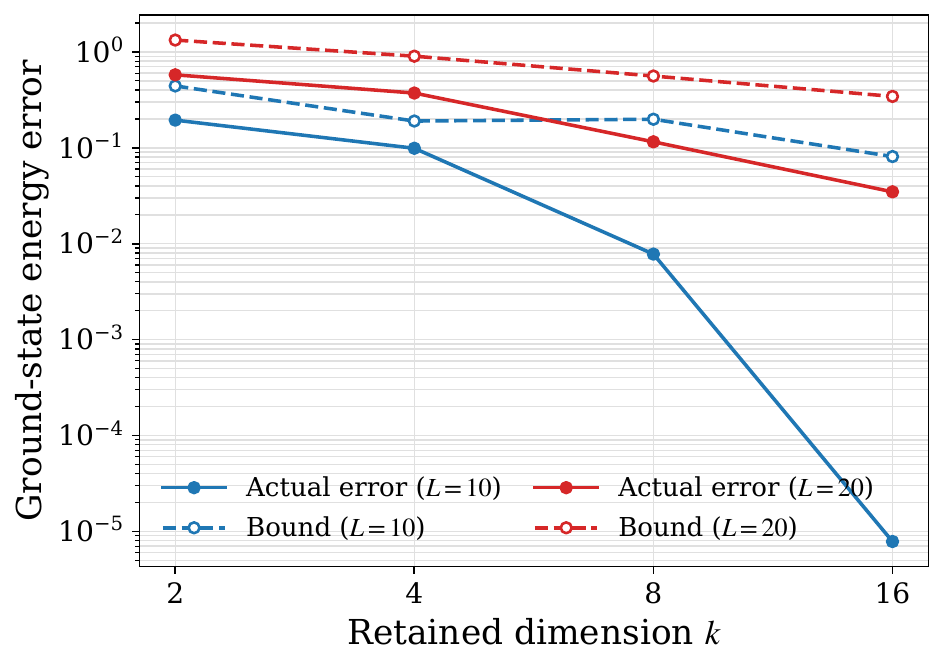}
\caption{
Numerical validation of the nonperturbative error bound for chains of lengths $L=10$ and $L=20$ at $J=1.0$ and $g=0.7$.
Solid curves with filled circles show the actual ground-state energy error
$a-E_0(H)$, while dashed curves with open circles show the corresponding bound
$r(b,\gamma)$ from Eq.~\eqref{eq_single_merge_bound}.
Blue and red curves denote $L=10$ and $L=20$, respectively.
For all tested values of $k$, $\gamma>0$ and
$a-E_0(H)\leq r(b,\gamma)$.
}
\label{fig:bound-validation}
\end{figure}

To verify this bound, we evaluate the complete retained sector $P$ and its discarded complement $Q=I-P$ for chains of lengths $L=10$ and $L=20$. Figure~\ref{fig:bound-validation} compares the actual ground-state energy error with the corresponding bound in Eq.~\eqref{eq_single_merge_bound}. For every tested $k$, the discarded-sector gap remains positive and the actual error lies below the theoretical bound.

We next propagate the single-merge bound through the homogeneous recursive hierarchy. Write $M=L/L_0=2^n$, so that the recursive construction consists of
$n=\log_2M$ merge steps, with
$\ell=1,2,\ldots,n$ labeling the $\ell$th merge. Because the system is homogeneous, at step $\ell$ there are $M/2^\ell$ equivalent merges in the physical chain. Let $\delta_\ell(k)$ be the ground-state energy increment. Applying Eq.~\eqref{eq_single_merge_bound} gives
\begin{equation}
0\leq
\delta_\ell(k)
\leq
r_\ell(k)
=
\frac{
\sqrt{\gamma_\ell^2+4b_\ell^2}
-
\gamma_\ell
}{2}
\leq
\frac{b_\ell^2}{\gamma_\ell}.
\label{eq_level_error_bound}
\end{equation}
The total ground-state error is obtained by summing the contributions from all merge steps,
\begin{equation}
0\leq
\Delta E_0(L,k)
=
\sum_{\ell=1}^{n}
\frac{M}{2^\ell}
\delta_\ell(k)
\leq
M
\sum_{\ell=1}^{n}
2^{-\ell}r_\ell(k).
\label{eq_recursive_error_bound}
\end{equation}
Suppose that the single-merge error is bounded by
$r_\ell(k)\leq R(k)$ throughout the hierarchy. Equation~\eqref{eq_recursive_error_bound} then gives
\begin{align}
\Delta E_0(L,k)
&\leq
MR(k)
\sum_{\ell=1}^{n}2^{-\ell}
=
(M-1)R(k),
\notag\\
\frac{\Delta E_0(L,k)}{L}
&\leq
\frac{R(k)}{L_0}
\left(
1-\frac{L_0}{L}
\right)
<
\frac{R(k)}{L_0}.
\label{eq_extensive_error_bound}
\end{align}
Thus, the total truncation error grows at most extensively with $L$. The resulting bound on the error density is determined by the magnitude of
$b_\ell^2/\gamma_\ell$ at each merge step and by the elementary module size $L_0$. The numerical data in Fig.~\ref{fig:error-scaling} approach a nearly size-independent error density, consistent with this analysis.

\begin{figure}[t]
\centering
\includegraphics[width=\linewidth]{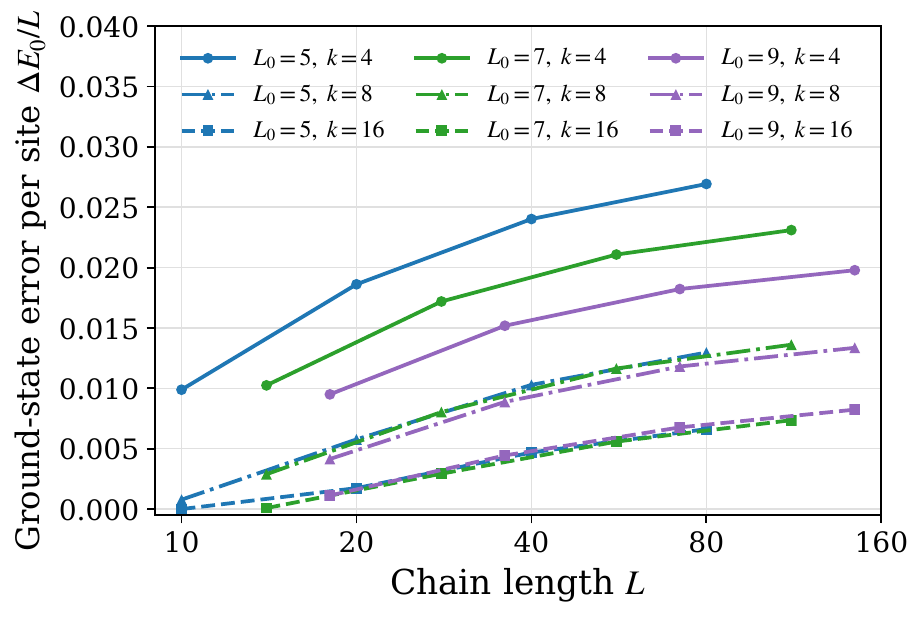}
\caption{
Ground-state error per site for elementary module lengths
$L_0=5$, $7$, and $9$ at $J=1.0$ and $g=0.7$.
Colors distinguish the elementary module sizes, while the line styles and markers distinguish the retained dimensions.
The reference ground-state energies are obtained from the exact free-fermion calculation.
}
\label{fig_L0_scaling}
\end{figure}

\begin{figure*}
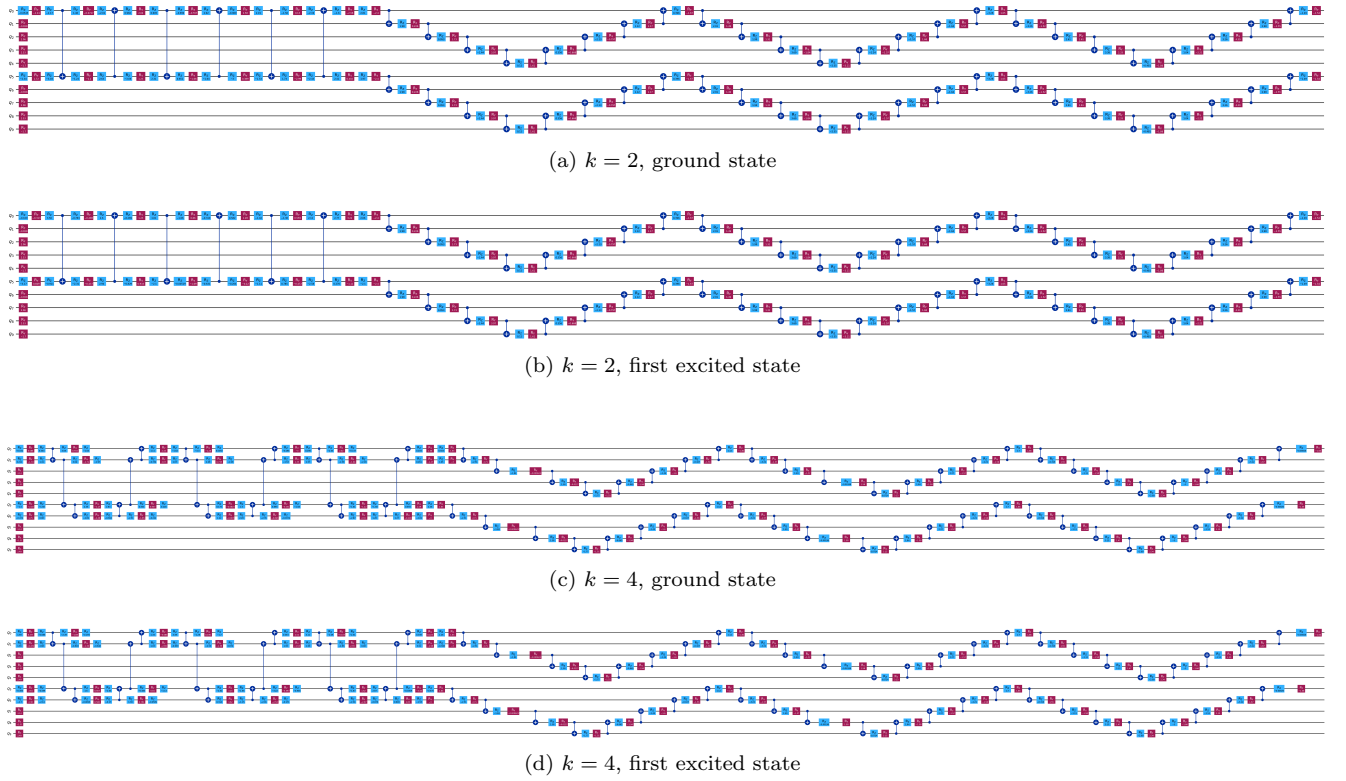

    \centering

    \subfloat[\(k=2\), ground state]{%
        \includegraphics[width=0.98\textwidth]{k2_3_3.png}%
    }\par

    \subfloat[\(k=2\), first excited state]{%
        \includegraphics[width=0.98\textwidth]{k2_3_3_1.png}%
    }\par

    \vspace{0.4cm}

    \subfloat[\(k=4\), ground state]{%
        \includegraphics[width=0.98\textwidth]{k4_3_3.png}%
    }\par

    \subfloat[\(k=4\), first excited state]{%
        \includegraphics[width=0.98\textwidth]{k4_3_3_1.png}%
    }\par

    \caption{Representative trained encoded circuits for the \(L=10\)
    calculation. The block encoders are trained with block depth \(3\),
    and the coupled system is optimized with logical depth \(3\).}
    \label{fig:trained-circuits-k2-k4}
\end{figure*}

The dependence on $k$ can be understood from the spectral weight removed at each merge. At merge step $\ell$, consider $H=H_A+H_B+\lambda V_\ell$
where $\lambda$ is introduced as a formal coupling parameter, and denote the product eigenstates of the two identical blocks by
$|\alpha\beta\rangle$. In a nondegenerate regime, the part of the second-order energy lowering omitted by retaining only $\alpha,\beta<k$ is
\begin{equation}
\delta E_{\mathrm{miss},\ell}^{(2)}(k)
=
\lambda^2
\sum_{\max(\alpha,\beta)\geq k}
\frac{
|\langle\alpha\beta|V_\ell|00\rangle|^2
}{
(E_\alpha^A-E_0^A)
+
(E_\beta^B-E_0^B)
}.
\label{eq_perturbative_tail}
\end{equation}
For the homogeneous system, we define the first discarded excitation scale at merge step $\ell$ as
\begin{align}
\Delta_{\ell,k}
&=
\min_{\max(\alpha,\beta)\geq k}
\left[
(E_\alpha^A-E_0^A)
+
(E_\beta^B-E_0^B)
\right]
\notag\\
&=
E_k^{(\ell-1)}
-
E_0^{(\ell-1)},
\label{eq_discarded_gap}
\end{align}
where the second equality follows because the two blocks are identical. Every denominator in Eq.~\eqref{eq_perturbative_tail} is therefore bounded from below by $\Delta_{\ell,k}$. Using completeness in the discarded subspace gives
\begin{align}
\delta E_{\mathrm{miss},\ell}^{(2)}(k)
&\leq
\frac{\lambda^2}{\Delta_{\ell,k}}
\sum_{\max(\alpha,\beta)\geq k}
|\langle\alpha\beta|V_\ell|00\rangle|^2
\notag\\
&=
\frac{\lambda^2}{\Delta_{\ell,k}}
\langle00|
V_\ell Q_k V_\ell
|00\rangle
\notag\\
&\leq
\frac{
\lambda^2\lVert V_\ell\rVert^2
}{
\Delta_{\ell,k}
},
\label{eq_perturbative_bound}
\end{align}
where $Q_k$ projects onto the discarded product states. Thus, increasing $k$ suppresses the omitted contribution by moving the discarded-sector cutoff to higher excitation energies and by including more of the boundary-coupled spectral weight in the retained subspace. Equivalently, it increases $\Delta_{\ell,k}$ and decreases
$\langle00|V_\ell Q_kV_\ell|00\rangle$ in Eq.~\eqref{eq_perturbative_bound}.

The scaling analysis also characterizes the trade-off between accuracy and computational cost. For fixed $L$ and $L_0$, we estimate the error bound as
$R(k)\simeq C_Rk^{-p}$,
where $C_R$ and $p$ are determined over the relevant finite range of $k$. According to Eq.~\eqref{eq_homogeneous_complexity}, the $k$-dependent total merge cost can be written as $T_k\simeq c_Tk^6$,
where $c_T$ includes the fixed factor $\log_2(L/L_0)$. Equation~\eqref{eq_extensive_error_bound} then gives
\begin{equation}
\frac{\Delta E_0(L,k)}{L}
\lesssim
\frac{C_R}{L_0}
\left(
1-\frac{L_0}{L}
\right)
\left(
\frac{T_k}{c_T}
\right)^{-p/6}
=
\mathcal{O}\!\left(T_k^{-p/6}\right).
\label{eq_cost_error_tradeoff_appendix}
\end{equation}
This is the finite-range cost--accuracy relation summarized in Eq.~\eqref{eq_cost_error_tradeoff} of the main text.

Finally, we examine whether the observed error scaling depends on the choice of the elementary module size by repeating the calculation for $L_0=5$, $7$, and $9$. As shown in Fig.~\ref{fig_L0_scaling}, increasing the length of the initial block reduces the error density, with the reduction being particularly pronounced for small $k$. However, changing $L_0$ does not alter the overall scaling behavior with $L$. For each fixed $L_0$ and $k$, the increase in the error density becomes progressively weaker as $L$ grows, consistent with an approach to a size-independent value, while increasing $k$ systematically reduces the error for every module size. These observations show that our qualitative analysis of the size dependence remains valid across different choices of $L_0$.

\section{Representative trained encoded circuits}
\label{app:trained-circuits}

The circuit-level formulation introduced in the main text uses block encoders to prepare the retained low-energy subspace of each module, followed by a logical variational circuit acting within the tensor-product space of the encoded modules. To illustrate this two-stage construction explicitly, this appendix presents representative trained circuits used in the $L=10$ calculations for $k=2$ and $k=4$, as shown in Fig.~\ref{fig:trained-circuits-k2-k4}. These circuits implement the same conceptual decomposition as the projected eigensolver. Here, local encoders define the truncated low-energy subspace, and the logical circuit explores the low-dimensional tensor-product space of the retained module states. The hardware-oriented formulation therefore provides a route to replacing the exact local diagonalization in the classical algorithm by variationally trained quantum encoders.

\bibliographystyle{apsrev4-2}
\bibliography{references_modular_encoded_subspace}

\end{document}